\documentclass[journal]{IEEEtran}

\usepackage{cite}
\usepackage{amsmath,amssymb,amsfonts}
\usepackage{algorithmic}
\usepackage{graphicx}
\usepackage{multirow}
\usepackage{textcomp}
\usepackage{xcolor}
\usepackage{caption}
\usepackage{subcaption}
\usepackage[belowskip=-5pt,aboveskip=0pt]{caption}
\usepackage{etoolbox}
\usepackage{booktabs}

\ifCLASSINFOpdf
\else
\fi
\IEEEoverridecommandlockouts\IEEEpubid{\makebox[\columnwidth]{978-0-7381-3207-5/21/\$31.00
~\copyright~2021 IEEE \hfill} \hspace{\columnsep}\makebox[\columnwidth]{ }}

\begin{document}

\title{DSM-MoC as Baseline: Reliability Assurance via Redundant Cellular Connectivity in Connected Cars}

\author{Emeka Obiodu,~\IEEEmembership{Member,~IEEE,}
       Aravindh Raman, 
        Abdullahi Abubakar,~\IEEEmembership{Student Member,~IEEE,}\\
        Simone Mangiante,~\IEEEmembership{Member,~IEEE,}
        Nishanth Sastry,~\IEEEmembership{Senior Member,~IEEE,}
        Hamid Aghvami,~\IEEEmembership{Fellow,~IEEE}

\thanks{E. Obiodu and H. Aghvami are with King's College London, UK. Email: chukwuemeka.obiodu@kcl.ac.uk and hamid.aghvami@kcl.ac.uk.
A. Abubakar and N. Sastry are with University of Surrey, UK. E-mail: a.abubakar@surrey.ac.uk and n.sastry@surrey.ac.uk.
A. Raman is with Telefonica I+D, Spain. Email: aravindh.raman@telefonica.com.
S. Mangiante is with Vodafone, UK. Email: simone.mangiante@vodafone.com.
}}

\newif
\ifcomment
\commenttrue

\ifcomment
\newcommand\ns[1]{\textbf{\textcolor{red}{NS: #1}}}
\newcommand\ar[1]{\textbf{\textcolor{blue}{AR: #1}}	}
\else
\newcommand\ns[1]{}
\newcommand\ar[1]{}
\fi

\maketitle

    \thispagestyle{empty}

\begin{abstract}

Connected Cars (CCs) and vehicle-to-everything (V2X) use cases require stringent reliability for safety and non-safety uses. With increasing network softwarisation, it has become easier to use multiple, redundant connectivity options instead of relying on a single network connectivity. But where should these redundant connections be managed? Is it at a network provider's core network - i.e. supply side managed (SSM) - or at the CC - i.e. demand side managed (DSM)? In our work, we investigate the use of SSM and DSM for CCs on four separate days and across 800 kilometers of major / minor roads in South East England. For Day 1, we captured performance indicators, and determined hypothetical multi-operator configurations for four UK providers and a global Universal SIM. For Day 2, 3 \& 4, we built and deployed a test-bed to actually implement network switching and understand performance (incl. for TCP \& UDP) either on the road or in a stationary location. Based on our results, we make three contributions. First, we show that DSM can deliver superior performance for CCs more than any individual network (up to 28 percentage points in a hypothetical scenario), or SSM which had up to 4.8x longer page load times. Second, unlike other smartphone-only studies, our system-level study demonstrates that improvements of at least 12\% can be obtained in a practical DSM field implementation for a CC. Third, we confirm that the advantage of DSM in a field implementation is higher for UDP traffic (23\% better latency) compared to TCP (13\%).

\end{abstract}

\begin{IEEEkeywords}
cellular, V2X, connected cars, multi-operator, DSM-MoC, SSM-MoC, redundancy, reliability, national roaming, universal SIM, global SIM, eCall.
\end{IEEEkeywords}

\IEEEpeerreviewmaketitle

\section{Introduction}

\subsection{Preliminaries}
\label{preliminaries}

Connected Cars (CCs) - using connectivity to communicate with other vehicles, the transport infrastructure, pedestrians or with the internet - are becoming mainstream. Analyses Mason notes that 164 million passenger cars were connected by the end of 2018 (about 16.5\% of the total) and forecast this will rise to 831 million by 2027 (about 56\% of total)\cite{mackenzie2019}. Cars use connectivity for safety and non-safety reasons, including infotainment, navigation and control. This is much broader than the cliched quest to support `driverless or autonomous cars'. In general, there are four groups of CC Vehicle-to-Everything (V2X) use cases: Vehicle-to-Vehicle (V2V), Vehicle-to-Person (V2P), Vehicle-to-Infrastructure (V2I) and Vehicle-to-Network (V2N)\cite{5g20155g}. The two competing CC connectivity technologies are the IEEE802.11p based V2X version - also known as Dedicated Short Range Communication (DSRC) - and Cellular V2X, although we note that the ecosystem seems to be coalescing on cV2X as the choice~\cite{nguyen2017comparison}. cV2X (the focus of this paper) has been happening with 2G, 3G and 4G technologies and the expectation is that 5G will provide improved reliability, especially to support safety-critical use cases\cite{obiodu20195g}.

Given their mobility, CCs endure significant geographical variation in connectivity performance as they move from one base station to the other. ~\cite{andrade2017connected} notes that this reality means that, compared to smartphones, there is only a small window to deliver large volumes of data, making it imperative to assure reliability in that small window. There are generally five approaches to smartly managing the network to improve and assure the reliability of CC connectivity: (i) the use of Sidelink, introduced in 3GPP Release 16, for near field communications\cite{molina2017lte}, (ii) improvements in the underlying reliability of the network to support V2N use cases (e.g. moving from 4G to 5G)\cite{li20145g}, (iii) adoption of prioritisation for different service classes (e.g. via 5G Network Slicing)\cite{campolo20175g}, (iv) use of bespoke/private networks on the roads\cite{lecklider2018autonomous} and (v) via multi-operator connectivity or national roaming agreements on public cellular networks\cite{husain2019ultra}. While (i) is ideal for V2V/V2P/V2I, the expectation is that (ii) to (v) will support V2N scenarios. Likewise, skepticism about QoS assurance from (ii) is driving developments in (iii) to (v).

\begin{figure*}[t]
\centering
\includegraphics[width=14cm, height=3.5cm]{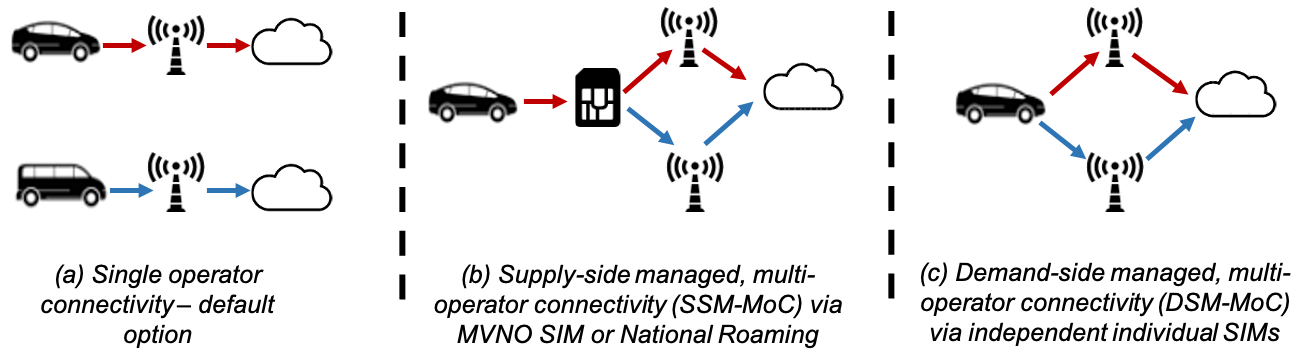}
\caption{Vehicle-to-Network (V2N) connectivity options for Connected Cars (CCs). While most existing CC contractual relationships with cellular network operators are based on Option (a), some MVNOs are stepping in to offer Option (b). But would Option (b) provide better reliability and QoS assurance than the hardly used Option (c)?}
\label{fig_cc_model}
\end{figure*}

\subsection{Motivation}
\label{motivation}

In this paper, we focus on (v) and compare three options for providing multi-operator CC connectivity (Figure~\ref{fig_cc_model}). Option 1 (Figure~\ref{fig_cc_model}a) is the default. CCs are connected via either an embedded or an aftermarket module, to a single public cellular network with an exclusive contractual relationship between the network provider and the car manufacturer or user. 

Users who demand better reliability, or where regulation compels it (e.g. e-Call in Europe~\cite{oorni2017vehicle}), can fall back on another network. This is Option 2 (Figure~\ref{fig_cc_model}b), a supply-side managed multi-operator connectivity (SSM-MoC) option where the CC connects to a single service provider who then manages the actual network connectivity in the backend. The service provider is either a network operator who relies on `national roaming' (i.e permitted to use other networks\cite{gligorijevic2017role}) or a mobile virtual network operator (MVNO) who relies on wholesale deals with multiple operators. While Option 2 looks like a good redundancy option, the decision on which network to use is still made far away from the user, by an entity whose priorities may not always align with that of the user. Accordingly, in Option 3 (Figure~\ref{fig_cc_model}c), the decision is repatriated to be managed in situ as a demand-side managed multi-operator connectivity (DSM-MoC) setup.

\subsection{The role of softwarization}
\label{softwarisation}

The opportunity to use redundant DSM-MoC is an example of how increasing softwarisation of the cellular infrastructure is opening up previously impossible usage scenarios. Softwarisation enables the integration of intelligence at the device, edge and core into a composite framework for improved operations \& management of networks. Unlike SSM-MoC which is a continuation of the `intelligent core, dumb node' philosophy in the operations of cellular networks, DSM-MoC assumes that each network node is imbued with intelligent decision-making capabilities to optimize network reliability. This explains the push to deploy edge computing - including Multi-access Edge Computing (MEC) - to run user applications closer to the user or to deploy/run virtualized core network functions - e.g. 5G User Plane Function (UPF) - at the edge\cite{porambage2018survey}.

We adopt this softwarised and intelligent edge/device philosophy in our work on DSM-MoC. Given the increasing computational power in modern cars, CCs are effectively edge computing nodes and can leverage the intersection of edge and device intelligence to smartly manage their own connectivity in situ. This is similar to the problem of access point selection in large, multi-provider Wi-Fi networks \cite{raschella2017quality}

\subsection{Focus of work}
\label{focus}

\begin{figure}[t]
\centering
\includegraphics[width=0.65\columnwidth]{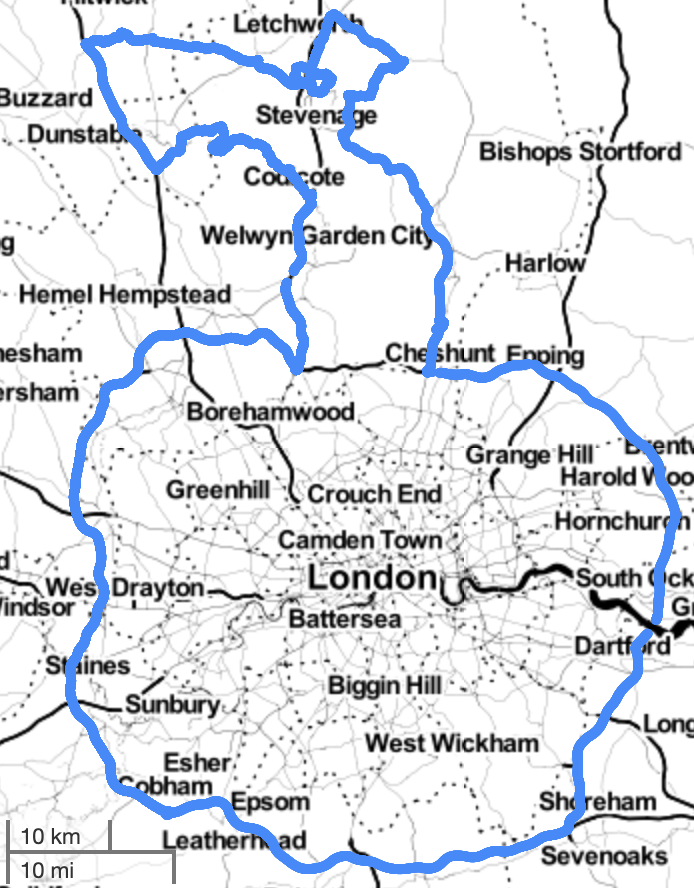}
\caption{Drive route during the measurement for Day 1 \& 2 as mapped by the GPS tracker on the Android application}
\label{fig:route_taken}
\end{figure}

To investigate the performance that can be seen for the three options, we conducted field measurements on a stretch of major and minor roads in the UK on four days: 15 November 2020 (Day 1), 14 June 2021 (Day 2), 15 November 2021 (Day 3) and 16 November 2021 (Day 4) - see Figure \ref{fig:route_taken}. Dates were selected for convenience as Day 1/2/3 measurements are self containing without the need for comparisons across dates. In total, we drove for $>$800km, across all our road tests. Day 1/2 drives included 192km on the M25, the UK's busiest motorway, and another 188km on rural roads in the Hertfordshire / Bedfordshire countryside. Day 3 was a 40 km drive on the A1(M) motorway in Hertfordshire while Day 4 was done at a suburban stationary location in Hertfordshire. 

\begin{figure*}[t]
\centering
\includegraphics[width=17.5cm, height=6.5cm]{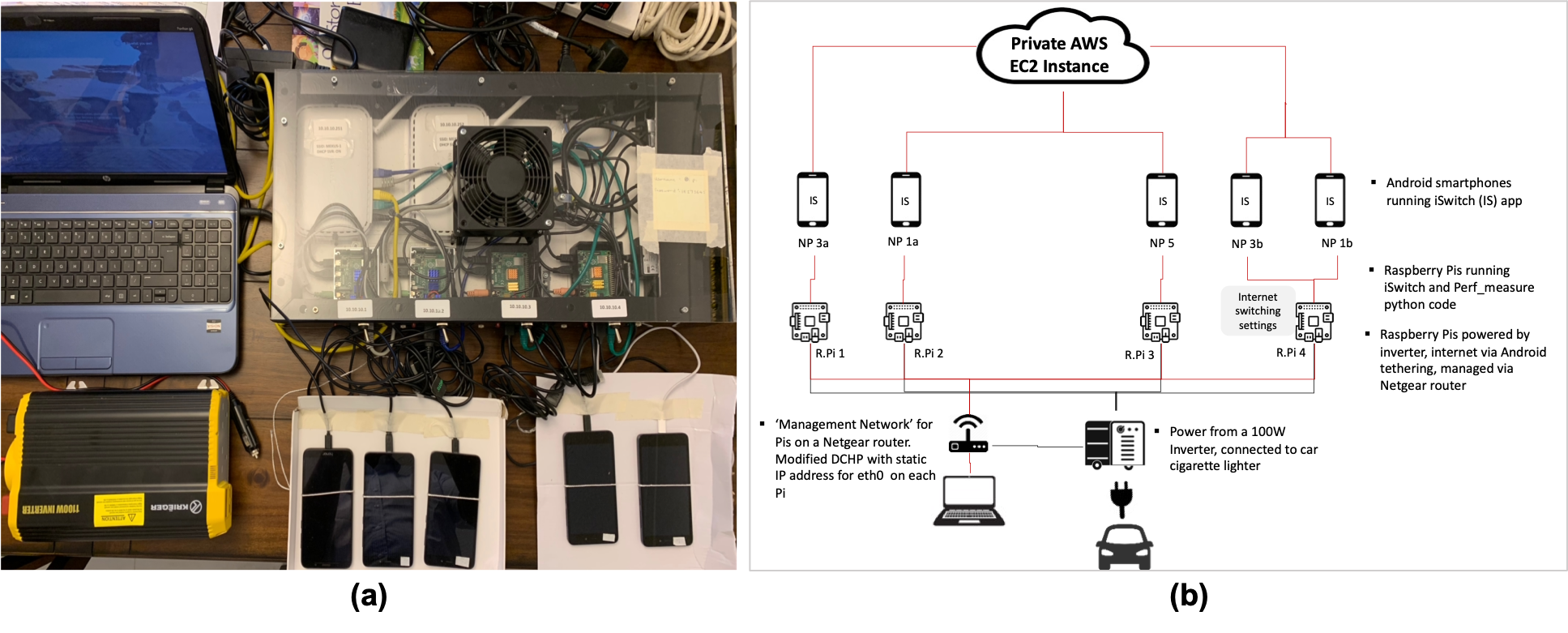} 
\caption{Experiment setup for field measurements on Day 2: (a) top view layout of equipment - laptop, inverter, five Android devices and a box with four Raspberry Pis and two netgear routers; (b) the schematic layout of the measurement setup}
\label{setup}
\end{figure*}

There are five hardware and four software components for our test bed (Figure \ref{setup}). The hardware is comprised of (i) five Xiaomi M1 4G devices, (ii) four Raspberry IV devices, (iii) two Netgear 10/100 ethernet routers, (iv) a HP laptop and (v) a 100 watts inverter. The (ii) and (iii) components are enclosed in a specially constructed glass-top box and the entire hardware setup is shown in Figure \ref{setup}a. We used only the five devices and the inverter for Day 1. The four specially-designed software components are: (i) CellPerf Android app for network parameter measurements on Day 1, (ii) iSwitch Android app as ``cellperf-lite'' that can monitor at extended granularity on Day 2, (iii) iSwitch python module running on the Raspberry Pis to switch networks based on latency and download a 1.3MB webpage on Day 2, (iv) client/server versions of tcpServer \& udpServer python modules to measure latency, jitter \& packet loss for different TCP/UDP packet sizes on Day 3 \& 4.

We designate the four UK network providers as Network Provider (NP) 1 - 4 and the universal SIM as the SSM-MoC or NP 5. In general, we avoid disclosing the identities of NP 1 - 5 in order to focus the analysis on the merits of redundant connectivity instead of comparisons of UK network providers. For Figure~\ref{fig_cc_model}, either NP 1 or 3 provides insight on Option 1, NP 5 is for Option 2 and the switching between NP 1 \& 3 is for Option 3. This work combines insights from Day 1, 2, 3 and Day 4. On Day 1, the focus was on coarse-grained measurements for NP 1 - 5 and preliminary results have been published\cite{obiodu2021cc}. On Day 2, we focus on fine-grained measurements for NP 1 and 3 as the two best in-country networks and NP 5. On Day 3, the focus was to understand the behaviour of different TCP and UDP packet windows during network switching in a high velocity scenario for NP 1, 3 and 5. Day 4 provides insights on the same TCP/UDP behaviour for NP 1, 3 and 5, but in a stationary environment.

\subsection{Novel contribution}
\label{novel}

Our work is about `reliability arbitrage' in a system-implementation on a CC, compared to other works on multi-connectivity (e.g. Google Fi, iCellular) which are mostly aimed at `price \& quality arbitrage' on smartphones\cite{braud2019multi, odikayor2012dual, li2016icellular}. By using a system-level implementation with onboard computers (i.e Raspberry Pi), networking equipment and all the inherent unknown factors in a practical, real-life CC scenario, this approach demonstrates the advantage of DSM-MoC in a real-life and real-time scenario.

The results of Day 1 have been preliminarily published in \cite{obiodu2021cc}. In it, we made two main contributions:
\begin{enumerate}
    \item We quantified, via extensive field measurements on live networks in the UK, that using a hypothetical DSM-MoC redundancy setup can deliver superior performance advantages over the best-performing single-operator option. This can be up to 28 percentage points for latency.
    \item We showed that a hypothetical edge-based DSM-MoC can offer significantly better outcomes than cloud-based SSM-MoC which only achieves up to 4.8x longer page load times. The emphasis on DSM is contrary to common practice today where, under the concept of `national roaming'\cite{gligorijevic2017role}, it is assumed to be the responsibility of the network provider to provide SSM redundancy. 
\end{enumerate}

For Day 2, 3 \& 4 measurements, we built \& deployed a system implementation of DSM-MoC on a CC on live UK networks.  Based on the Day 2/3/4 measurements, we make additional three contributions:

\begin{enumerate}
    \item We demonstrate that despite the challenges in switching networks in real time in a deployment scenario, DSM-MoC still outperforms both SSM-MoC and single network options. Our results show that the median page load times for DSM-MoC 1/3 of 103 milliseconds was 12\% better than for NP 3, 382\% better than SSM/NP 5 and 1332\% better than NP 1. This validates the hypothetical results from Day 1 and provides a compelling need for commercial and regulatory action to encourage use of DSM-MoC for safety-critical systems.
    \item We show that DSM-MoC delivers greater improvements to UDP compared to TCP. For RTT, this was 23\% for UDP compared to TCP's 13\%. For jitter, this was 67\% for UDP compared to 47\% for TCP. This has practical implications for safety critical services in CCs that rely on UDP protocol such as voice/video communications, computer gaming, and the emerging trend to create 3D simulation and virtual worlds for the metaverse. 
    \item We demonstrate that DSM-MoC benefits are applicable at the typical packet sizes used for data communication (i.e. ~1500 bytes) with measurements using 1024 bytes achieving the best outcomes compared to 200 bytes or 2048 bytes. This observation is helpful because it means there is no need to re-architect the internet infrastructure and plumbing to support DSM-MoC.

\end{enumerate}

\subsection{Paper organisation}
\label{novel}
The remaining of this paper is organized as follows. Section \ref{sec:ecosystem} provides the context/background to connectivity for CCs while Section \ref{sec:reliability} introduces reliability theory and the use of redundancy in complex systems. In Section \ref{sec:system}, we introduce the system model and derive the theoretical reliability parameters that inform our work. Section \ref{sec:experiment} lays out how our measurement was setup, describing the design, hardware, software and logistics for our field experiment. In Section \ref{sec:result}, we present the results from Day 1, 2, 3 \& 4 measurements and discuss the implications for the CC ecosystem in Section \ref{discussion}. Finally, we provide a conclusion in Section \ref{sec:conclusion}.

\section{Setting the Context}
\label{sec:ecosystem}

\subsection{Vision for Connected Cars}
Today, CCs use cellular connectivity for real-time navigational support with maps, in-car infotainment (incl. in-car WiFi hotspots), updating firmware over-the-air (FOTA), remote car diagnostics and support (e.g. by the car manufacturer), monitoring of driving habits (e.g. by insurance companies) and for emergency response support (e.g.\ eCall in Europe~\cite{oorni2017vehicle}). In futuristic models of society, the vision is that all cars are connected, creating an ecosystem of cars, transport infrastructure and ancillary maintenance, operational and infotainment services\cite{obiodu20195g} and a market for data that is generated by CCs. The 5G-PPP whitepaper~\cite{5g20155g} provides the joint vision of the automotive and telecommunication industries. This vision has informed the activities of the 5G Automotive Association (5GAA), bringing together automotive manufacturers and ICT providers to develop and support interoperable and reliable cellular V2X services. Based on developments since 3GPP Release 14~\cite{wang2017overview}, the major new capabilities envisioned for CCs in the future are autonomous driving and inter-connectivity with other cars, pedestrians and road infrastructure. While these are most suited for Sidelink, current market deployments are still predominantly reliant on V2N connectivity for both safety and non-safety use cases.

\begin{table*}[t]
\caption{3GPP V2X categories and main KPIs (Adapted from \cite{campolo20175g})}
\label{tab:v2x_cat}
\begin{center}
\begin{tabular}{|p{1.3cm}|p{4cm}|p{2cm}|p{1.6cm}|p{4.3cm}|p{1.8cm}|}

\hline
Focus & V2X category & Comms type & Latency & Throughput & Reliability\\
\hline
Safety & Safety in driving & V2V, V2P, V2I & 100 ms & Not a concern & Not yet explicit\\
\hline
Safety & Traffic efficiency (esp. Platooning) & V2V, V2P, V2I & 100 ms & Not a concern & Not yet explicit\\
\hline
Safety & Tele-operated driving & V2N & 20 ms & 25 Mbps uplink for video \& sensors data, 1 Mbps for downlink & 99.999\%\\
\hline
Safety & Advanced driving & V2V, V2I, V2N & 1 ms & 10 Mbps for uplink and downlink & Nearly 100\%\\
\hline
Non-Safety & Car internet and infotainment & V2N & 100 ms & 0.5 Mbps for browsing, up to 15 Mbps for video & Not a concern\\
\hline
Non-Safety & Remote diagnostics and support & V2I, V2N & Not a concern & Not a concern & Not a concern\\
\hline
\end{tabular}
\end{center}

\end{table*}

\subsection{Performance expectations for Connected Cars}
To realise the vision of an integrated and reliable ecosystem around CCs, there are performance expectations for safety and non-safety V2X scenarios (Table~\ref{tab:v2x_cat}). Many of the current use cases are for non-safety scenarios and will remain so in the future. These include high data rate infotainment, in-car WiFi hotspots, navigational map updates, remote diagnostics and support. ~\cite{campolo20175g} note, however, that 3GPP is focused on four safety-related V2X scenarios:

\begin{enumerate}
    \item Safety: Focusing on the use of extended sensors to gather and exchange information among cars, road infrastructure and pedestrians. The aim is to improve safety and reduce the number of road traffic accidents.  
    
    \item Traffic efficiency: Covering platooning to enable a selection of cars to dynamically form groups, positioning themselves very close to each other, and moving together at high speeds as one unit. This mimics the railways and the aim is to increase road network capacity.  
    
     \item Tele-operated driving: For scenarios where a remote driver or V2X Application Server can remotely take control of a car (e.g. for dangerous driving conditions).
     
    \item Advanced driving: Covering both semi-autonomous and fully-autonomous driving, and is facilitated via the exchange of data between cars and road infrastructure. This is the \emph{holy grail} of the future of driving.
    
\end{enumerate}

\subsection{State of market for Connected Cars}
As the number of CCs in the field grows, it becomes relevant to evaluate if the visions and expectations for the CC ecosystem are being met. Continued growth in the number of CCs confirms that the vision is broadly supported. So we focus on two outcomes: scaling the CC ecosystem and achieving expected cellular network performance for CCs.

On scaling, the original vision for scaling CCs assumed that companies in the telecommunication and automotive industries will be the primary providers, developers and gate keepers for CC connectivity and solutions\cite{5g20155g}. In practice, since 2014, Apple's CarPlay and Google's Android Auto have become the dominant solution for infotainment services on CCs\cite{ramnath2020interacting}. Today, in most countries, Apple and Google have disintermediated car manufacturers to become the primary gatekeepers of non-safety services for CCs and the car owner/driver brings along their existing smartphone connectivity. 

On performance, \cite{grapeup2020} provides an extensive examination of the state of the CC ecosystem, noting the need to collaboratively address some long held assumptions to ensure that cellular connectivity remains relevant over the typical 12 to 15 years design lifecycle of the car industry. They caution that any chosen cellular connectivity option ``must deliver reliable, seamless, uninterrupted coverage in all countries and markets where the vehicles are sold and driven". This view is supported by \cite{mckinsey2021} who noted that many stakeholders in the CC ecosystem work in isolation on hard-to-scale, island and exclusive relationship between two players.

\subsection{Overcoming challenges to cellular connectivity}

Given the challenges of \emph{performance} above, we argue that the current setup of cellular connectivity for CCs is insufficient to deliver full benefits for the CC ecosystem (the challenge of scaling is mostly of commercial and environmental importance and so is out of scope). For performance for safety cV2X scenarios, CCs today experience the same performance as other network users and understanding this is a well trodden research area. For example, for 3G, ~\cite{paul2011understanding} showed that there is significant temporal and spatial variations in throughput in different parts of a 3G data network, albeit with predictable aggregate behavior. For 4G, and focusing on usage in atypical scenarios, ~\cite{erman2013understanding} studied performance during the Superbowl, finding causes and triggers for performance unpredictability. For 5G, \cite{obiodu20215g} notes that early 5G networks did not outperform 4G at the busiest five train stations in the UK. 

Given the variability in network performance, several authors have proposed technical, commercial and policy remedies to assure reliability that can benefit CCs. ~\cite{baltrunas2014measuring} observed that the reliability of the cellular network is much lower than imagined but that this can be improved significantly if devices can multi-home (i.e network sharing). ~\cite{petrov2018achieving} explores how to achieve end-to-end reliability of mission-critical traffic in 5G networks. ~\cite{campolo20175g} propose a V2X 5G network slice. ~\cite{labriji2021mobility} propose to optimize content placement in Multi-access Edge Computing (MEC) for the Internet of Vehicles. ~\cite{obiodu2019clasp} pushes for a regulatory  mandate for a 999-style (or 112/911) prioritisation that can be used for CC control signals.

\subsection{Comparing V2X and smartphones}

Although CCs and smartphones experience the same network conditions today, there are still differences in how they respond to those conditions. Using data from one million CCs in the USA (from a single manufacturer) and over one billion cellular radio connections, ~\cite{andrade2017connected} suggests that CCs are different to smartphones in three ways: 

\begin{enumerate}
    \item Comparison with smartphones: Connectivity for CCs and smartphones exhibit similar weekly and diurnal patterns. However, as CCs spend less time connected (overall and to each base station), the window to deliver large volumes of data is small and hence require different management approaches (e.g. for FOTA updates).  
    
    \item Comparison to IoT devices: Connectivity for CCs and IoT devices similarly only use few carriers, connect only to a subset of the network, and spend only a short time on the network and per session.   
    
    \item Unique traits: CCs are unique in the way they connect to different cells on different days, and predictably, the connectivity pattern matches commute-time patterns. Accordingly, ISPs could use prediction models to efficiently deliver content, manage mobility and assure QoS during the frequent handovers under high speed.  
\end{enumerate}

\subsection{Improving reliability of V2X}

Given the different remedies that have been proposed to improve reliability, and taking cognizance of the peculiarities of CCs vs smartphone, we summarise five broad approaches to improving and assuring the reliability of cV2X connectivity. 

\begin{itemize}
    \item Sidelink: Introduced in 3GPP Release 16 to support V2V, V2P and V2I use cases such as maintaining safety while driving, interacting with roadside infrastructure, avoiding accidents with pedestrians and for platooning. It uses the dedicated 5.9 GHz spectrum for Intelligent Transport Systems (ITS), to allow CCs to communicate directly with one another or other nearly entities, without the need for the wider, cellular network\cite{molina2017lte}.
    
    \item Generic generational improvements: Going from 4G to 5G will provide a performance boost to connectivity to support many safety and non-safety V2N use cases\cite{li20145g}. 
    
    \item Prioritisation: This is recommended for use when routine network upgrades are insufficient for reliability/security. 4G QoS Class Identifier (QCI), 5G QoS Indicator (5QI) and 5G Network Slicing are the key mechanisms for delivering prioritisation for V2X use cases\cite{campolo20175g} and 3GPP has already prescribed QCI/5QI 75/79 for them.
    
    \item Private 4G/5G networks: For scenarios where there is no public cellular network (e.g. supporting Rio Tinto's Autonomous Haulage Systems in Australian mines\cite{lecklider2018autonomous}) or where customers prefer to own/manage the communications infrastructure (e.g as seen in the long term vision articulated by Highways England on the possibility of building its own 5G network for UK motorways\cite{highways2017uk}).
    
    \item Multi-operator connectivity: Given the inability of any single network provider to provide QoS assurance, some customers may prefer to embrace multi-operator connectivity\cite{husain2019ultra}, either by using multiple redundant SIM cards, using an MVNO provider/Universal SIM card with multi-operator wholesale partners (e.g. Transatel or Twilio) or via roaming agreements (e.g. as several car makers are doing in Europe)\cite{blackman2018roaming}. This is the focus of our work and we provide a background in Section \ref{sec:reliability}.
\end{itemize}

\section{Redundancy Analysis for Connectivity}
\label{sec:reliability}

\subsection{Reliability \& Resilience in Complex Systems}
Reliability and resilience are fundamental design considerations for all complex systems in biology, computer science, ecology, economics, environmental science, engineering etc\cite{fraccascia2018resilience}. \cite{clark2016s} clarifies that reliability is the ability of a system to perform as designed while resilience is the ability of a system to bounce back after a perturbation. Reliability is a key network KPI for CCs (Table ~\ref{tab:v2x_cat}) and there are two completely different expectations on it. The first is from the original ITU-R ``Minimum requirements related to technical performance for IMT-2020 radio interface(s)" which specifies a minimum reliability requirement of  1-10\textsuperscript{-5} success probability of transmitting a layer 2/3 packet within a required maximum time~\cite{sg052017draft}. Operationally, it is rare to see any cellular network being defined by this. Instead, system reliability is typically the user-experienced reliability for a sporadic end-user. 

For a CC, geographical variation in connectivity performance is a major concern given that in the course of its journey, a CC will be handed over from one base station to the other. ~\cite{andrade2017connected} observes that this operational reality leaves only a short time window for large data transfers. For systems using the default single-path TCP, disruptions in the single connectivity path is a major source of performance degradation and has informed the development of Multi-path TCP as a means of providing redundant multi connectivity paths \cite{raiciu2012hard}.

\subsection{Redundancy for reliability \& resilience}

Generally, redundancy is the tried-and-tested means of improving reliability in biological and man-made engineered systems. Within engineering, the use of redundancy to assure reliability and resilience is an established approach and is applicable across all layers of the stack in the designed system\cite{sterbenz2014redundancy}. \cite{downer2009failure} argues that ``redundancy is the single most important engineering tool for designing, implementing and proving reliability in all complex, safety-critical technologies". Best practice in design is to ensure that the redundant units are independent of each other\cite{popov2003estimating}, providing diversity or distributed redundancy akin to the `degeneracy' seen in biological systems\cite{randles2011distributed}. Using the example of the crash of the European Space Agency (ESA) Arianne 5 rocket on 4 June 1996, \cite{downer2009failure} illustrates the importance of `design diversity' to achieve independence. For most ultra-reliable computer infrastructure (e.g datacentres), redundant paths and multi-homing are now the default operational setup. As `datacentres on wheels' (Intel's justification for spending \$15bn in 2017 to acquire Mobileye), we argue in this paper that redundant connectivity paths ought to be the default setup for CCs to support safety-related use cases.

\subsection{Redundancy for performance-constrained reliability}

While national roaming or SSM-MoC should theoretically improve resilience, it is unclear whether a switchover will happen for under-performance (but not outright failure) on one or several performance indicators (cf: train companies in the UK will provide a replacement bus service when the train line is closed but will not do so if trains are only running with a delay). We suggest that SSM is a continuation of the `Always Best Connected' concept\cite{gustafsson2003always} of the 3G era where cellular network operators expected to decide how users are connected to cellular or wi-fi. Yet, over the past 15 years, the trend has been unmistakable in the opposite direction with users (or user-controlled devices) making the decision.

Network users can make decisions based on any of the constituent attributes of system reliability or cost - e.g. use of multi-SIM devices \cite{odikayor2012dual} and Software defined wide area network (SD-WAN) solutions\cite{luciani2019mpls}. ~\cite{baltrunas2014measuring} explains that these reliability attributes include network reliability (e.g. failures, availability, radio conditions), data plane reliability (e.g. packet loss, loss runs, large events) and performance reliability (e.g. latency, HTTP throughput, SIP success rate). Classic reliability theory focuses mostly on network reliability and uses `availability' as the KPI of interest to ascertain if the system is functioning or not: i.e a system is functioning if, and only if ($\mbox{if{f}}$), a connectivity path exists between the input and output~\cite{romeu2004understanding}. But a connectivity path can be available yet grossly unable to deliver the performance required - necessitating path dependent reliability analysis ~\cite{aggarwal1982capacity}. 

\subsection{Redundancy setup for cellular services}
The use of multi-SIM devices is fairly common in developing countries as a mechanism for price and quality arbitrage\cite{odikayor2012dual}. In most developed countries, a multi-SIM approach has typically been regarded as unnecessary until Google launched Google Fi in April 2015 in the US. However, \cite{li2016icellular} note that Google Fi does not deliver effective multi-access switching. Instead, they propose iCellular as a client-side service to let commodity mobile devices customize their own cellular network access. In reality, despite initial market excitement, momentum for Google Fi has waned and the service, or similar, has not been launched outside the US because it offers little to relatively affluent customers in countries with decent cellular coverage. Our work on multi-connectivity for CCs focuses on a use case that has safety (and as such regulatory) consequences instead of only price comparison.  

Operationally, the use of embedded SIMs (e-SIM)\cite{gsma2018esim} and blockchain-based Accountable Just-in-time (AJIT) smart contracts\cite{faisal2020ajit} will make DSM-MoC easier. But there will be new problems in the way data packets are handled. Studies such as \cite{xiao2014tcp} have already shown that TCP behaves poorly in high-mobility environments. They note that, because of larger RTT and jitter, TCP throughput in high speed rail is 3x worse than static and 2x worse than driving scenarios. Swapping networks brings additional complication. \cite{li2018measurement} notes that multipath TCP improves performance across cellular providers while \cite{xu2020first} explored how different TCP variants improved performance over 4G in a high speed, multi provider scenario.

\subsection{Global feasibility of redundant cellular connectivity}
In practice, achieving independent parallel redundant cellular connections should be mostly feasible across the world because, apart from Djibouti, Eritrea \& North Korea, all countries have at least two operational cellular infrastructure. However, most publicly confirmed pilots or commercial contracts between a cellular network operator and a provider of safety-critical system (e.g. public safety LTE for emergency services\cite{doumi2013lte} such as AT\&T's Firstnet in the US or Testra's LANES in Australia) have touted the exclusivity of the relationship, with at best supply-side managed fall back in exceptional circumstances. Such supply-side managed redundancy is akin to hospitals (e.g. under the UK's Health Technical Memorandum 06-01\cite{health2017electric}) expecting their utility provider to also provide the legally-mandated standby power generator.

We explore experimentally the experience of redundant cellular connectivity, managed at the demand side.

\section{System Model}

\label{sec:system}

Given a CC connectivity system, with several connectivity options, which multi-operator connectivity option will experimentally deliver superior overall system reliability? We explore this problem statement theoretically for two dimensions: first, the acceptable threshold required for each reliability metric; second, the number of multi-operator connectivity options needed to optimise system reliability. 

\subsection{Threshold for reliability parameters}
In a system with instantaneous Network Reliability $R\textsubscript{i}$ that is based on the system's availability, we define the instantaneous Performance-constrained Reliability function $Q\textsubscript{i}$ as the reliability when the overall system performance meets a pre-determined threshold. In other words, $R\textsubscript{i}$ represents the reliability when the system is available; $Q\textsubscript{i}$ represents the reliability when the system is available and meets the required performance threshold. Therefore,

\begin{equation}
    Q\textsubscript{i} = W\textsubscript{i}  R\textsubscript{i}  \label{eq}
\end{equation}

and $W\textsubscript{i}$ is an integer binary function [0, 1]. 
$ W_i = 0 [W_i < 1]$; 
$ W_i = 1 [W_i \geq 1]$.

There are both endogenous and exogenous levers to optimize $R\textsubscript{i}$ for a cellular connectivity system. Upgrading from 3G to 4G to 5G is the most critical endogenous lever to improve intrinsic reliability.  Beyond that, exogenous levers become important to manage customer demand, and mitigate sub-optimal network outcomes. Here, we focus on the two main exogenous levers: probability of network congestion ($Y\textsubscript{i}$) and the probability of network connectivity availability ($Z\textsubscript{i}$.)

In order words, if the instantaneous cumulative failure rate $\lambda$\textsubscript{i} of the system is
\begin{equation}
    \lambda_i = \int_{0}^{\infty} (Y\textsubscript{i} +  Z\textsubscript{i})dt
\end{equation}

then the instantaneous reliability R\textsubscript{i} is

\begin{equation}
    R\textsubscript{i} = 1 - \lambda_i = 1 - \int_{0}^{\infty} (Y\textsubscript{i} +  Z\textsubscript{i})dt
\end{equation}

We can explore $Y\textsubscript{i}$ and $Z\textsubscript{i}$ using mathematical modeling. For the probability of network congestion $Y\textsubscript{i}$, we use the Hazard Rate model as a form of Survival Analysis to approximate the probability that a particular connection is congested. The probability density function of the Hazard Rate model is:

\begin{equation}
    Y\textsubscript{i} = t \phi e\textsuperscript{-p}
\end{equation}

where \emph{t} is the instantaneous time during operations, $\phi$ represents the $>$0 hazard rate per inherent failure of the system, and \emph{p} (-1 $<$ p $<$ 1) is the changing rate of the network traffic density.

For the probability of network connectivity availability $Z\textsubscript{i}$, we use the Weibull distribution to approximate the probability that a particular network connection is available. The probability density function of the Weibull distribution is given by: 

\begin{equation}
    Z_i = \frac{\beta}{\eta}\Bigl(\frac{t}{\beta}\Bigr) ^{\beta - 1}  e^{-(\frac{t}{\eta})^\beta}  
\end{equation}

where $\beta$ and $\eta$ are the shape and scale parameters respectively and are $>$ 0.

The Hazard Rate and Weibull Distribution probability density functions provide a theoretical framework to investigate the failure rate ($\lambda_i$) and reliability ($R\textsubscript{i}$) of each individual network connection for CC.

\subsection{No of multi-operator connectivity options needed}

In a CC system with $n$ parallel redundant networks, the overall system reliability can be approximated as a composite function of the Hazard Rate and Weibull Distribution of each individual network. If each of the $n$ parallel redundant networks has a uniform failure rate $\lambda$\textsubscript{i}, then:

\begin{equation}
    R_i = 1 - \Bigl(1 - e^{-\lambda_i} \Bigr)^n
\end{equation}

and the Mean Time To Fail (MTTF) $\mu$ is

\begin{equation}
    MTTF = \mu = \int_{0}^{\infty} R\textsubscript{i} dt
\end{equation}

In the UK with four cellular networks, each with same failure rate $\lambda$\textsubscript{i}, the instantaneous $R$\textsubscript{i} are: 

\begin{equation}
    \begin{aligned}
        & n_1 = e^{-\lambda} \\
        & n_2 = 2e^{-\lambda} - e^{-2\lambda} \\
        & n_3 = 3e^{-\lambda} - 3e^{-2\lambda} + e^{-3\lambda}\\
        & n_4 = 4e^{-\lambda} - 6e^{-2\lambda} + 4e^{-3\lambda} - + e^{-4\lambda}\\
    \end{aligned}
\end{equation}

and the MTTF $\mu$\textsubscript{i} are

\begin{equation}
    \begin{aligned}
        & n_1 = \frac{1}{\lambda} \\
        & n_2 = \frac{3}{2\lambda} \\
        & n_3 = \frac{11}{6\lambda}\\
        & n_4 = \frac{25}{12\lambda}\\
    \end{aligned}
\end{equation}

\begin{figure}
     \centering
     \begin{subfigure}[b]{0.45\columnwidth}
         \centering
         \includegraphics[width=\textwidth]{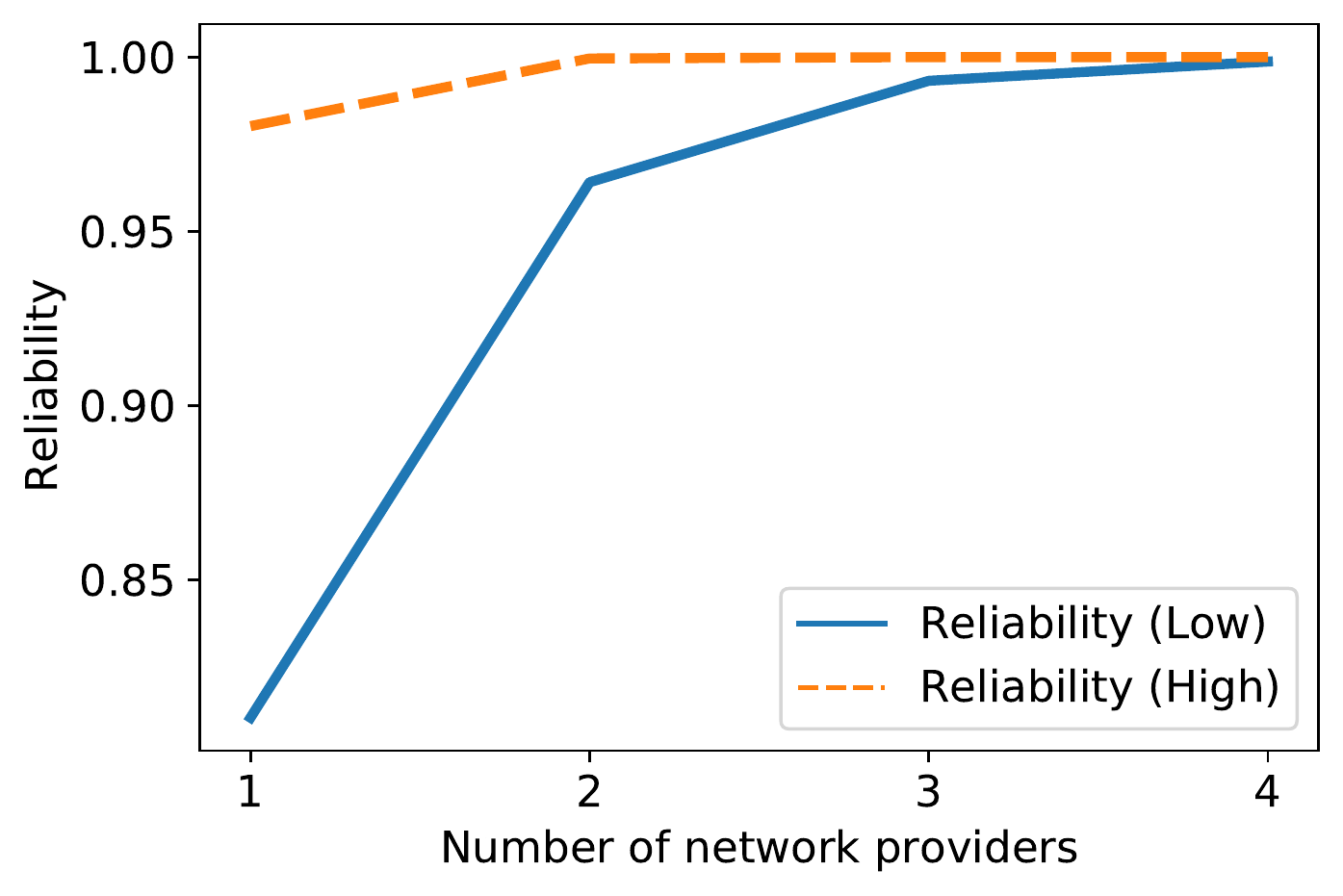}
         \caption{Reliability}
         \label{fig:reliability_theory}
     \end{subfigure}
     \hfill
     \begin{subfigure}[b]{0.45\columnwidth}
         \centering
         \includegraphics[width=\textwidth]{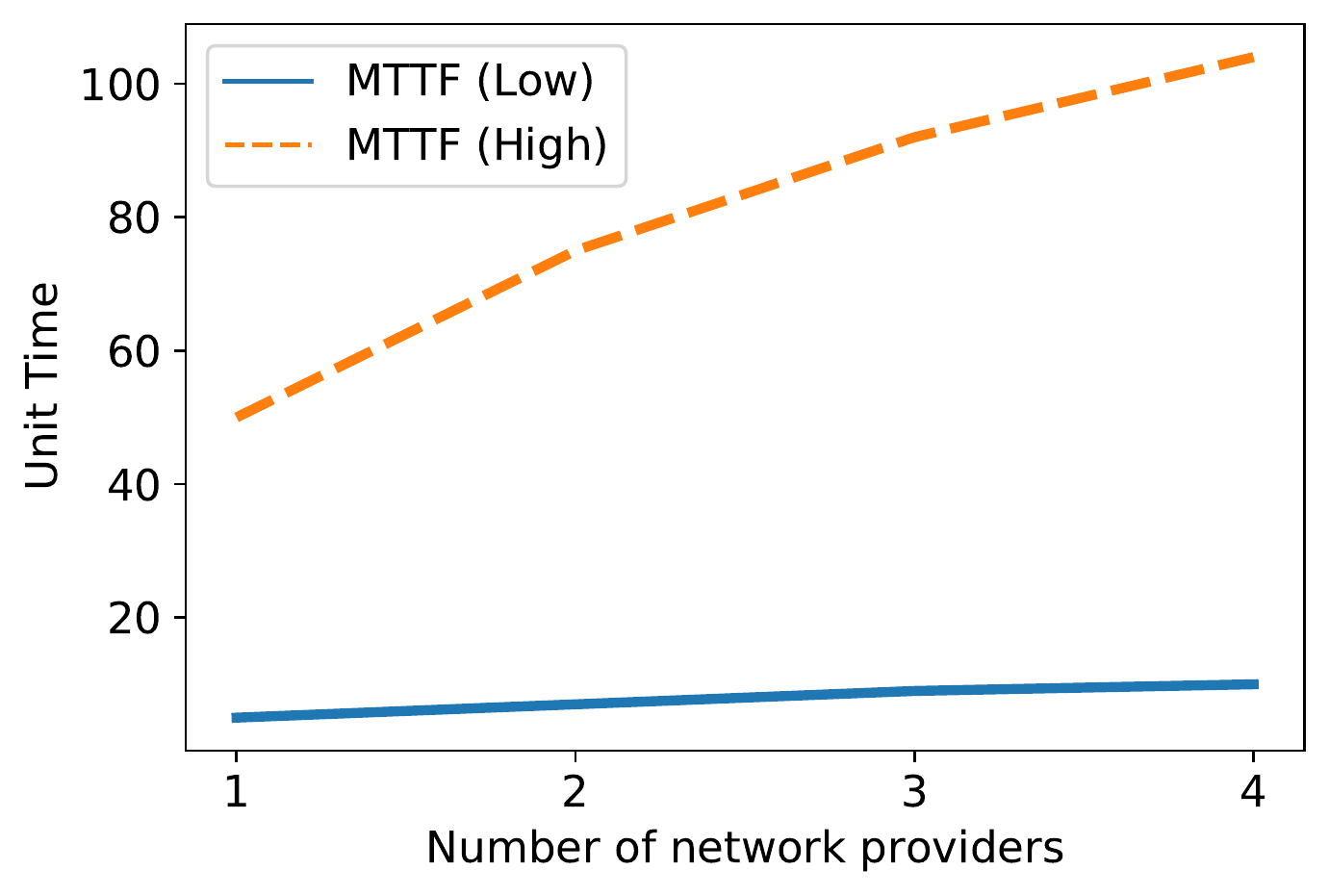}
         \caption{MTTF}
         \label{fig:mttf_theory}
     \end{subfigure}
     \vspace{2mm}
    \caption{Theoretical evaluation of expected Reliability and Mean Time to Fail based on 4G penetration levels from  \protect\cite{ofcom2020connected}}
        \label{rmf_theory}
\end{figure}

In its Connected Nations 2020 report\cite{ofcom2020connected}, UK telecoms regulator, Ofcom, reports that UK operators provide outdoor coverage to 98\%-99\% of premises and their networks’ coverage of the UK landmass ranges from around 79\% to around 85\%. We use 98\% as the upper limit and 79\% as the lower limit for the expected availability (i.e unavailability / failure rate: 2\% - 21\%). Figure \ref{rmf_theory} summarises the aggregate $R$\textsubscript{i} and $\mu$\textsubscript{i} expected for a road test across South East England. The field measurement is designed to confirm and validate these.

\section{Experiment design, parameters \& setup}
\label{sec:experiment}

\subsection{Experiment Philosophy}

Our work is informed by the observations in the three preceding chapters: the challenges facing cellular connectivity for CCs (Section \ref{sec:ecosystem}), the exposition about the role of redundancy as a mitigating option for unreliability of a single network connection (Section \ref{sec:reliability}) and the theoretical evaluation of how redundancy can improve overall system reliability by lowering the probability of network congestion and network failure (Section \ref{sec:system}). These inform our hypothesis that a redundant connectivity setup will deliver higher values for $R\textsubscript{i}$ and $Q\textsubscript{i}$. 

\subsection{Experiment Goals \& Design}
We had four key experiments \& design goals:

\begin{enumerate}
    \item Experiment 1: Quantify the value of the hypothetical advantage of using a redundant connectivity setup. This investigates the hypothesis in Section \ref{sec:ecosystem} that individual 3G, 4G and early 5G networks are unable to meet the stringent performance requirements for CCs. For this, we measured 4G performance for all four UK providers. 
    \item Experiment 2: Determine the best setup for redundant connectivity options (i.e. SSM-MoC vs DSM-MoC). This explores the practicalities of designing redundant systems for a given engineering problem as described in Section \ref{sec:reliability}. For this, we measured 4G performance on a global Universal SIM provider in the UK and compared it with the performance of the four UK providers.
    \item Experiment 3: Assuming DSM-MoC is hypothetically better, explore setting up actual DMS-MoC implementation to validate hypothetical result. This explores the implementability of redundant systems (as per Section \ref{sec:reliability}) and investigates the impact of network congestion and failures on DSM-MoC (as per Section \ref{sec:system}). For this, we built a system implementation of DSM-MoC. 
    \item Experiment 4: Investigate suitability of different types of applications for DSM-MoC and compare performance in stationary and high-speed scenarios. Given the differences between smartphones and CCs (Section \ref{sec:ecosystem}), this explores to what extent the cellular connectivity can serve CCs in different usage scenarios(Section \ref{sec:system}). For this, we used our system implementation to explore behavior of TCP \& UDP for DSM-MoC.
\end{enumerate}

\subsection{Experiment Route and dates}
The experiments were conducted on major and minor roads, and at a suburban location in South East England. This was done on four dates that were selected based on convenience as each of the experiments was self contained and there was no need for cross-day comparisons. Our road drive was on a stretch of roads in South East UK totaling 237 miles (380km) twice on 15 November 2020 (Day 1) and 14 June 2021 (Day 2), and then 28 miles (44.8km) on 15 November 2021 (Day 3). The stationary measurements were at a suburban location in Stevenage, on 16 November 2021 (Day 4). Including other exploratory drives for testing, we drove for over 1,000km. 

Day 1 \& 2 drives included most of the length of the M25, the 120 miles (192km) ring road around London - the busiest motorway in the UK - and another 117 miles (188km) drive through rural roads in Hertfordshire and Bedfordshire. Day 3 was on the A1(M). The justification for measuring on both major and minor roads is to establish the connectivity experience for a CC on all road types. This is because any expectation of a future of pervasive connectivity for CCs will need to consider major roads (assumed to be adequately covered) and minor rural roads with low traffic density. This is summarised in Table \ref{tab:experiment}.

\subsection{Hardware Setup}
\subsubsection{Android Devices} 
We used five Xiaomi M1 4i devices (released April 2015). The CPU of the devices is the Octa-core (4x1.7 GHz Cortex-A53 \& 4x1.0 GHz Cortex-A53), running on the Qualcomm MSM8939 Snapdragon 615 (28 nm) chipset. The devices support Android SDK version of up to API level 23. For Day 1, these were each connected to SIM cards for Operators NP 1 - 4 and SSM-MoC (NP 5). For Day 2, 3 \& 4, there was no NP 2 and NP 4. Instead, there were two NP 1 and NP 3 connections. While NP 1 - 4 SIM cards were picked up locally and the cost of data usage on them was approximately £1 / GB, NP 5 SIM card had to be specially ordered and its aggregate cost of data was £95 / GB.

\subsubsection{Raspberry Pis}
For Day 2, 3 \& 4, we used four Raspberry Pi 4 devices (released June 2019). Each of the devices has the Broadcom BCM2711 SoC processor with a 1.5 GHz 64-bit quad-core ARM Cortex-A72 processor, and 1 MB shared L2 cache; had 8GB of RAM and ran the Raspbian 10 Buster operating system. They were housed in our specially-constructed glass-top box and are connected to the internet via USB tethering from the Android devices. Three of the Pis were connected to NP 1, 3 and 5. The fourth was connected to both NP 1 and 3 with the ability to switch between them. 

\subsubsection{Management Network}
For Day 2, 3 \& 4, to access the Pis without individual screens, keyboards and mice, we setup a management network made up of a HP Pavilion laptop (Windows 10, AMD E2-1800 processor and 6GB RAM) and two Netgear DG834G routers with four ports, port speeds of 10/100 and maximum LAN speeds of 100Mbps. The routers are housed in the same box as the Pis and we used two routers because we needed five ethernet ports (four Pis and the laptop) but each router had only four ports. We modified DCHP with static IP address for eth0  on each Pi and then from the laptop, we used SSH and Microsoft Remote Desktop to access the Pis. Microsoft Remote Desktop was used for the initial setup and during the actual field measurement while SSH was faster for copying out results of our measurements from the Pis.

\subsection{Software Setup}
\subsubsection{Measurement software - CellPerf}
\label{cellperf}

For Day 1, we used the specially designed \emph{Cell\_Perf} Android measurement app, which is based on an adaptation of Multiping-for-Android app\cite{Softgearko2015}. CellPerf enables us to conduct 24 different measurements for 14 websites every 5 minutes. These include RTT on both TCP port 80 and TCP echo port 7, PLT, packet loss, jitter, uplink speed, downlink speed, network type (e.g. LTE, HSPA+ etc), base station ID, base station location, device CPU parameters, device RAM usage. The choice of websites was based on their Alexa Ranking as at December 2019 and on perceived importance to the UK digital society. In total, we recorded 728 readings per device in 5 hours 20 minutes, a total of 3,640 measurement cycles and 87,360 individual readings.

\begin{table}[t]
\caption{Summary of experiments 1 - 4}
\label{tab:experiment}
\begin{center}
\begin{tabular}{|p{1.7cm}|p{2.0cm}|p{4.0cm}|}
\hline
Experiment & Date & Route  \\
\hline
Experiment 1 & Day 1 (15 Nov 2020) & Done along 237 miles of major and minor roads in South East England  \\
\hline
Experiment 2 & Day 1 (15 Nov 2021) & Same route as Experiment 1 \\
\hline
Experiment 3 & Day 2 (14 Jun 2021) & Same route as Experiment 1 \& 2 \\
\hline
Experiment 4 & Day 3 (15 Nov 2021) \& Day 4 (16 Nov 2021) & Done along 28 miles of A1(M) motorway on Day 3 and at a stationary location in Stevenage on Day 4 \\
\hline

\end{tabular}
\end{center}

\end{table}

\subsubsection{Measurement \& Internet Switching software - iSwitch}
\label{iSwitch}
For Day 2, 3 \& 4, we used iSwitch.py to switch between two internet connections based on which one had better configuration. The iSwitch tool performed three roles. Firstly, it did a ping to our own AWS server every 3 seconds via the tethered phones to establish the latency. Secondly, based on an average of all measurements in every 10 seconds, the code switched between the available networks. Thirdly, the code measures page load times every 5 seconds by downloading a 1.3MB webpage. In total, 7.2K ping packets/phone were sent in 5 hours 37 minutes and the location and webpage download performance were monitored every 5 seconds.

\subsubsection{TCP/UDP observation software: tcpServer \& udpServer}
For Day 3 \& 4, we used separate client and server versions of tcpServer.py and udpServer.py to measure the performance of TCP and UDP for single network operations and for a DSM-MoC. We ran eight experiments concurrently each time (4 TCP and 4 UDP) and setup eight AWS EC2 Instances for each of the experiments. TCP experiments used port 4545 while UDP used 4646. We ran the experiments with different packet sizes in bytes (200, 1024, 2048 and 4096). For each iteration, we sent 10 packets, every minute, from the client (Raspberry Pi) to the server (AWS EC2 instance) and measured the time it took to receive all the packets at the server. The session timed out in 60 seconds if there was no confirmation from the server.

\subsection{Benchmarks for Analysis}
In table \ref{tab:benchmark}, we summarise the benchmarks for evaluating the coarse-grained measurements. These benchmarks are partly based on what is already obtainable on operational 4G networks globally (as reported by Opensignal)\cite{boyland2019}, plus the connectivity expectations for V2X use cases as stated in \cite{campolo20175g}.

\subsection{Comparison with Drive-Through Tests}
\label{drive_through}
There is a long and well-established history of drive through tests to measure cellular network performance. These are used by regulators, network providers, equipment vendors and other agencies to map, assess and improve network performance\cite{kora2016accurate}. However, a common methodological challenge for these is that most drive test systems are architecturally diverse, with incompatible interfaces and data formats\cite{riihijarvi2018machine}. We note also that these drive test systems are usually done only for limited number of times, in tightly controlled test environments, and with equipment that is carefully positioned and calibrated. In real life, a CC that needs to make decisions on connectivity options on a regular basis will have to rely on continuous measurement while using less-ideal equipment. Our methodology is designed to provide insights for this latter scenario.

\section{Results}
\label{sec:result}

We structure our results into three subsections. Firstly, we summarise the coarse-grained results from Day 1 (already published in \cite{obiodu2021cc}) and position it as the baseline for comparing Day 2 and Day 3/4 results. Secondly, we explore the fine-grained results from Day 2 based on actual switching of network connectivity in real-time. Finally, we analyse the relative impact on TCP and UDP based on Day 3/4 results.

\begin{table}[t]
\caption{Performance thresholds for Day 1 measurements}
\label{tab:benchmark}
\begin{center}
\begin{tabular}{|p{3.5cm}|p{3.5cm}|}

\hline
Metric & Benchmark  \\
\hline
Uplink speed & 25 Mbps \\
\hline
Downlink speed & 50 Mbps \\
\hline
Packet Loss & 0\% \\
\hline
Round Trip Times (RTT) & 100 ms \\
\hline
Jitter & 20 ms \\
\hline
Page Load Times (PLT) & 1000 ms \\
\hline
Availability & RTT $>$ 10,000 ms \\
\hline
\end{tabular}
\end{center}

\end{table}

\begin{figure*}{}
\centering
\includegraphics[width=15cm, height=4.5cm]{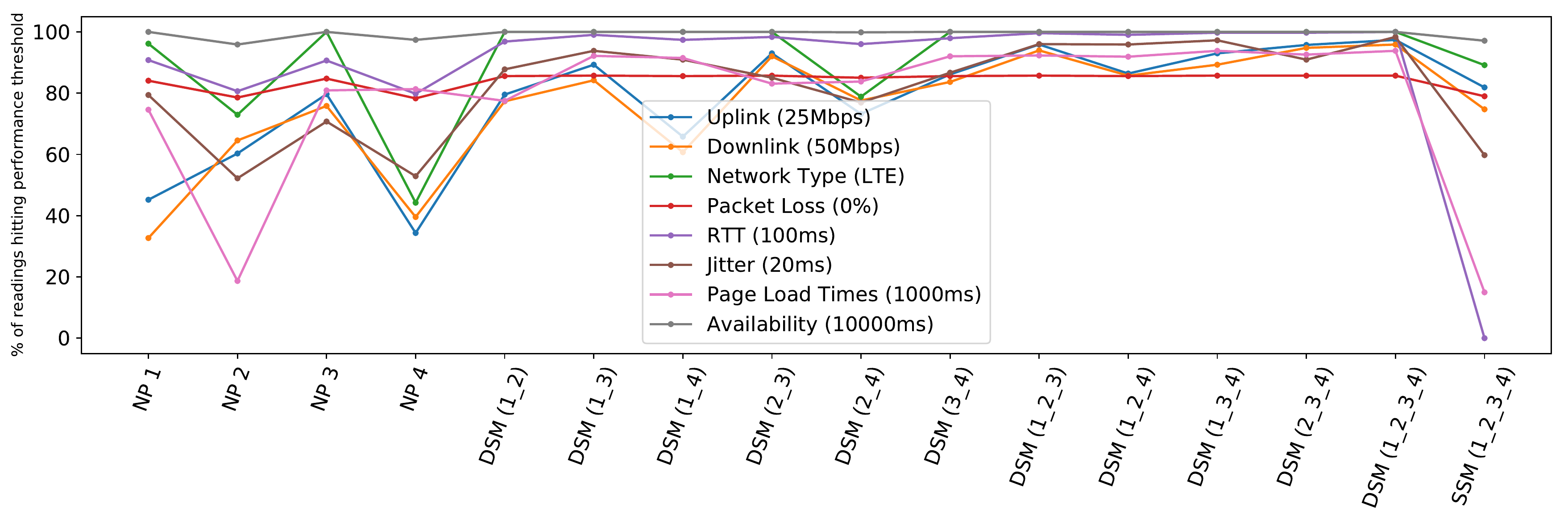}
\caption{Day 1 - Performance comparison for 8 measurement parameters: actual performance from 4 network providers (NP 1- 4) \&  Supply-Side Managed provider (SSM); projected performance for 11 combinations of Demand-Side Managed (DSM) setups. The clear out-performance of DSM shows that a multi-operator setup will provide better reliability and QoS assurance}
\label{fig:redundancy}
\end{figure*}

\subsection{Day 1 - Establishing the Hypothetical Benefit of DSM-MoC} 
\label{overall}

\subsubsection{DSM-MoC offers best hypothetical reliability}

In comparing the actual performance of NP 1 - 4, NP 5 and 11 hypothetical combinations of DSM-MoC, we show that DSM-MoC offers the best $Q\textsubscript{s}$ outcome. For NP 1 - 4 and NP 5 (SSM-MoC), $Q\textsubscript{s}$ is based on the field measurement on \emph{each network}. For DSM-MoC options, $Q\textsubscript{s}$ is based on meeting the threshold on \emph{any network}. Figure \ref{fig:redundancy} provides the overall picture from Day 1 measurements showing that there are significant performance gaps across all the four NPs  and SSM-MoC on all benchmarks. This makes it clear that any CC relying on any single NP is unlikely to have consistent reliability across time and in all locations. Out of the 3,640 measurements, 68\% achieved RTT of 100ms, 54\% for PLT of 1000ms, 60\% for uplink speeds of at least 25Mbps, 57\% for downlink speeds of at least 50Mbps and 80\% were connected to LTE. The cumulative performance data masks the diversity in performance on each of the NPs too; e.g. percentage of NP 2 PLT measurements $<$1000ms was only 19\% vs 75\% for NP 1 and 81\% for NP 3/4.

Hypothetically, it is evident from Figure \ref{fig:redundancy} that a DSM-MoC setup that is able to switch between NPs  will always provide a much higher reliability than any single-operator option or the SSM-MoC option. We show that if a four NP, DSM-MoC setup (i.e DSM-MoC 1/2/3/4) is implemented, overall system $Q\textsubscript{s}$ for RTT will improve to 100\%. For jitter, $Q\textsubscript{s}$ improves to 98\%, for uplink speed 97\%, for downlink speed 96\%, for PLT 94\%, and for packet loss 86\%. The delta between DSM-MoC 1/2/3/4 versus the best single-operator NP 3 is as high as 28 percentage points (pp) for jitter, 20pp for downlink speed, 18pp for uplink speed and 9pp for RTT.

\subsubsection{SSM-MoC delivers inferior app performance}

Although NP 5 (SSM-MoC) achieved the fastest median speeds of 25.4Mbps for uplink and 50.7Mbps for downlink compared to NP 1 - 4, its performance on packet loss, latency, jitter, packet loss and page load times were inferior, negating any benefits of the faster speed it achieves. For downlink/uplink speeds, \emph{getLinkDownstreamBandwidthKbps()} and \emph{getLinkUpstreamBandwidthKbps()} return the speeds in kbps. For latency (RTT recorded in accessing a TCP port 80 or TCP echo port 7 for particular websites from our app), the median RTT for NP 5 was 152ms, four times worse than NP 3's 38ms. For jitter (taking the average of five RTT readings), NP 5 was 15ms, same as for NP 3 and higher than NP 1's 11ms. For packet loss (observing packet losses recorded during our ping measurements), NP 5 is the lossiest path, losing 100\% of the packets in 18.3\% of readings. For page load times (using Android's webview), NP 5 again had the worst outcome with only 15\% of its readings meeting the 1000ms benchmark. These observations point to a profound conclusion: for a service that costs $>$95 times more, NP 5's performance is grossly inferior and does not warrant the cost.

\subsubsection{Availability benchmark is insufficient for CCs}

\begin{figure}[t]
\centering
\includegraphics[width=6cm, height=3.5cm]{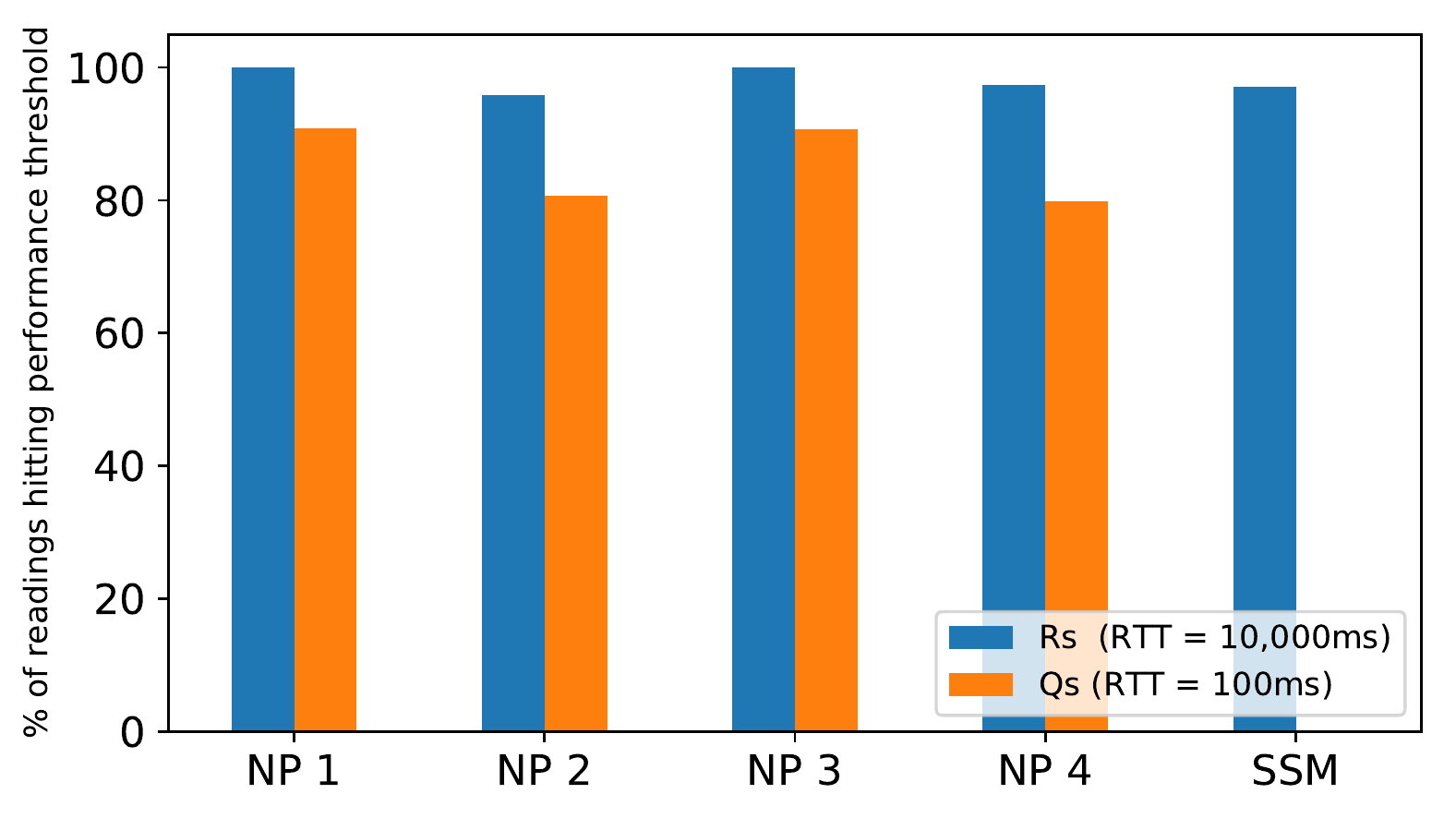}
\caption{Day 1 - Availability: Percentage measurements under 10,000ms vs 100ms. Benchmarks based on availability \emph{R(s)} differ from performance constrained reliability \emph{Q(s)}.}
\label{fig:availability}
\end{figure}

 Figure \ref{fig:availability} shows that for measurements on all five NPs, $R\textsubscript{s}$ is significantly better than $Q\textsubscript{s}$, providing confirmation that assessing reliability of CC connectivity based on availability will always show a better outcome than an assessment based on a performance constrained benchmark (i.e $R\textsubscript{s}$ $>>$ $Q\textsubscript{s}$). Using latency for analysis, we assess availability ($R\textsubscript{s}$) as an RTT of 10,000ms (10 seconds) - the Internet Control Message Protocol (ICMP) sessions default timeout - and a delay after which user experience, even for web browsing on smartphones, becomes unbearable\cite{mirkovic2018survey}. For $Q\textsubscript{s}$, we use 100ms which is required for competitive gaming\cite{mirkovic2018survey}, and is close to double the overall measured median RTT on NP 1 - 4. Accordingly, any official KPI or SLA that is benchmarked against availability $R\textsubscript{s}$ will suggest a better performance than a KPI based on  $Q\textsubscript{s}$.

\subsubsection{SSM-MoC performs poorly vs Ofcom data}

\begin{figure}[t]
\centering
\includegraphics[width=6cm, height=3.5cm]{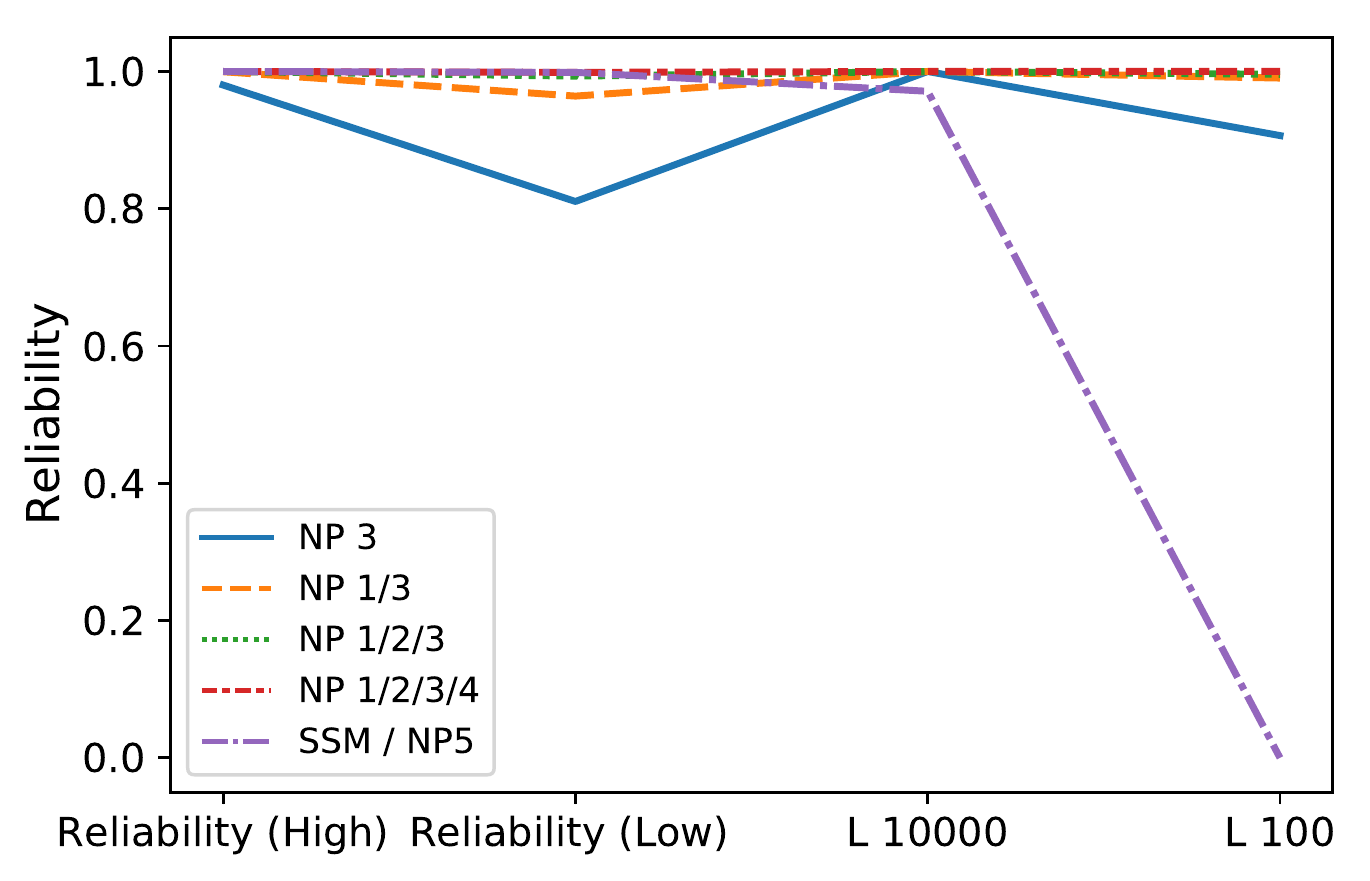}
\caption{Day 1 - Comparison with predicted reliability based on Ofcom data. It is clear that a 4-network SSM-MoC is inadequate, whether for availability (RTT = 10000ms; \emph{R(s)} = 97\%) or specific threshold (RTT = 100ms; \emph{Q(s)} = 0\%).}
\label{fig:cf_ofcom}
\end{figure}

Using Day 1 results, we compare the hypothetical reliability for the best combinations of DSM-MoC and SSM-MoC with the expected lower and upper reliability limits from Ofcom's coverage data.  For a 4-network redundancy system, Figure \ref{fig:cf_ofcom} shows that SSM-MoC performed significantly poor compared to the expectation. Using Availability (RTT = 10000ms), SSM-MoC achieves only 97\% even when NP 3 achieved 100\%. For RTT = 100ms, SSM-MoC was 0\% while NP 3 was 91\%. For the rest of the configurations, Figure \ref{fig:cf_ofcom} shows that the achieved reliability fits neatly between the lower and upper reliability limits from Ofcom data.

\subsection{Day 2: Implementing DSM-MoC \& Validating its Benefit}

\subsubsection{DSM-MoC achieves better performance overall}

\begin{figure}[t]
\centering
\includegraphics[width=6cm, height=3.5cm]{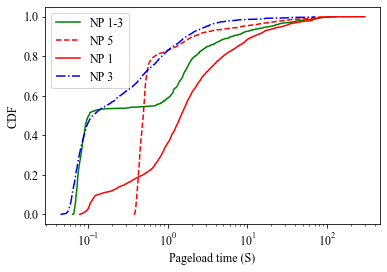}
\caption{Day 2 - Distribution of page load latency to a webpage of size 1.3MB hosted in AWS. CDF shows that, overall, the combined NP 1/3 DSM-MoC achieved a better page load times outcome than the SSM-MoC / NP 5 or either NP 1 or NP 3 }
\label{fig:pltcompare}
\end{figure}

As expected, NP 1/3 DSM-MoC delivered better overall performance than NP 5 (SSM-MoC) or using either of NP 1 or NP 3. We measured RTT every 3 seconds on each of the Raspberry Pis with iSwitch and took the average of the readings every 10 seconds to determine best network to switch to. For Pi 1- 3 with only a single network connection, there is no switching to do. For Pi 4, we switched between NP 1/3.  We measure actual application performance by measuring every 5 seconds on the Pis, the PLT for a 1.3MB webpage hosted in AWS. Figure \ref{fig:pltcompare} shows that the DSM-MoC of NP 1/3 achieved a median page load times of 103 milliseconds, compared to 496 for SSM-MoC, 115 for NP 3 and 1475 for NP 1; that is, DSM-MoC was 12\% better than NP 3, 382\% better than SSM/NP 5 and 1332\% better than NP 1. In doing so, \emph{we validate that DSM-MoC will deliver better outcomes whether in a hypothetical scenario (Day 1) or in reality (Day 2).}  

The emphasis on `overall' is because DSM-MoC did not consistently deliver the better outcome always. In fact, Pi 3 / NP 3 achieved the lowest maximum and lowest minimum page load times. This can be seen in Figure \ref{fig:pltcompare} too as the NP 3 began to consistently outperform the combined NP 1/3 option beyond the median mark when NP 1/3 began to track NP 1 more closely. This observation highlights the technical and operational challenges in implementing DSM-MoC where switching delays or unexpected hardware/software behaviours force a deviation from the anticipated performance.

\subsubsection{Predictive algorithms better for DSM-MoC}
While our setup is based on RTT readings every 3 seconds, and switching based on the average of readings every 10 seconds, our extensive evaluation of the latency readings in Table \ref{tab:reactive} show that there is a trade-off to make in terms of performance improvements vs number of switching required. In this case, switching based on a 10 seconds window will deliver 47.71\% improvement vs NP 1 and 5.95\% improvement vs NP 3 alone. But this requires 182 switches, creating more room for error and switching delays. This observation highlights the challenge to meeting key performance targets with SSM-MoC. For time critical scenarios, the delays in getting to the cloud-based switching centre and the in situ switching delays all contribute to degrading the realised performance outcomes.

Rather than using static switching windows, the DSM-MoC lends itself to using predictive tools to make the switching decision with a view to optimising either for performance or for number of switches. We can already see that a 60 second window achieves a not so different outcome from the 10 second window but with only 54 required switches. In this case, our hypothetical Oracle achieves the optimal performance improvements for NP 1 (58.13\%) and NP 3 (5.17\%), and also the fewest number of switches (52).

\subsubsection{Geographical variation in DSM-MoC performance}

\begin{figure}[t]
\centering
\includegraphics[width=0.65\columnwidth]{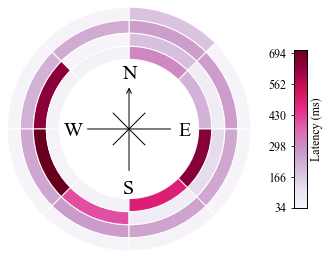}
\caption{Day 2: Latency comparison across different directions on our route for the four Raspberry Pis: Oracle (NP 1/3), NP 5, NP 3, NP 1 (from outer to inner)}
\label{fig_route_taken}
\end{figure}

For a CC, the challenges of implementing DSM-MoC can also be seen in the spatial variation of the latency readings along our 380 km route. In Figure \ref{fig_route_taken}, we show the latency readings along a North-East-South-West coordinate map, highlighting which of the four Raspberry Pis delivered the best latency readings. For the combined NP 1/3 Pi, we compared the latency that would have been recorded if our Oracle was taking the two latency readings from NP 1 and 3. It is evident that despite the spatial variations along the route, the combined setup is well positioned to realise the lowest latency readings.

\begin{table}[t]
    \centering
\begin{tabular}{lrrrr}
\toprule
  t (s) &          NP 1 (\%) &        NP 3 (\%)  & \#switches\\
\hline
 10 &  47.71 &  5.95 & 182\\
 20 &  47.62 &  6.18 & 132\\
 30 &  47.17 &  5.01 & 110 \\
  40 &  45.64 &  2.32 & 92\\
 50 &  46.21 &  3.07 & 64 \\
  60 &  47.08 &  5.17 & 54\\
  \hline
Oracle & 58.13 & 22.49  & 52\\
\hline
\end{tabular}
    \caption{Day 2: Latency readings per `t' seconds used for switching vs performance improvements \& no of switches. A DSM-MoC oracle will deliver a more optimal outcome}
    \label{tab:reactive}
\end{table}

\subsection{Day 3/4: Determining Impact of DSM-MoC on TCP/UDP}

\subsubsection{DSM-MoC benefits exist across different TCP/UDP scenarios}

Having confirmed from Day 2 measurements that DSM-MoC achieves better performance, we set out to understand the behavior of TCP/UDP as networks are switched. We did this for different TCP/UDP packet sizes and for a stationary vs mobile environment. Figure \ref{fig_tcp_udp} shows that, compared to a single operator, DSM-MoC                     achieves better performance for different TCP/UDP packet sizes and in both stationary and mobile environments. For simplicity, we compared with only NP 3 which has already proven to have the best network performance on the same route from Day 1 results.  From the 2165 readings recorded, we observe wide variation in the minimum and maximum RTT and jitter values for different TCP/UDP packet sizes. Overall, the median improvements were 13\% for RTT on TCP, 23\% for RTT on UDP, 47\% for jitter on TCP \& 67\% for jitter on UDP. Clearly, UDP benefited more from sharing, suggesting that for applications that rely more on UDP, a DSM-MoC approach is advantageous.

\subsubsection{DSM-MoC benefits were higher for stationary vs mobile scenario}
\begin{figure*}
     \centering
     \begin{subfigure}[b]{0.45\textwidth}
         \centering
         \includegraphics[width=6cm, height=4.0cm]{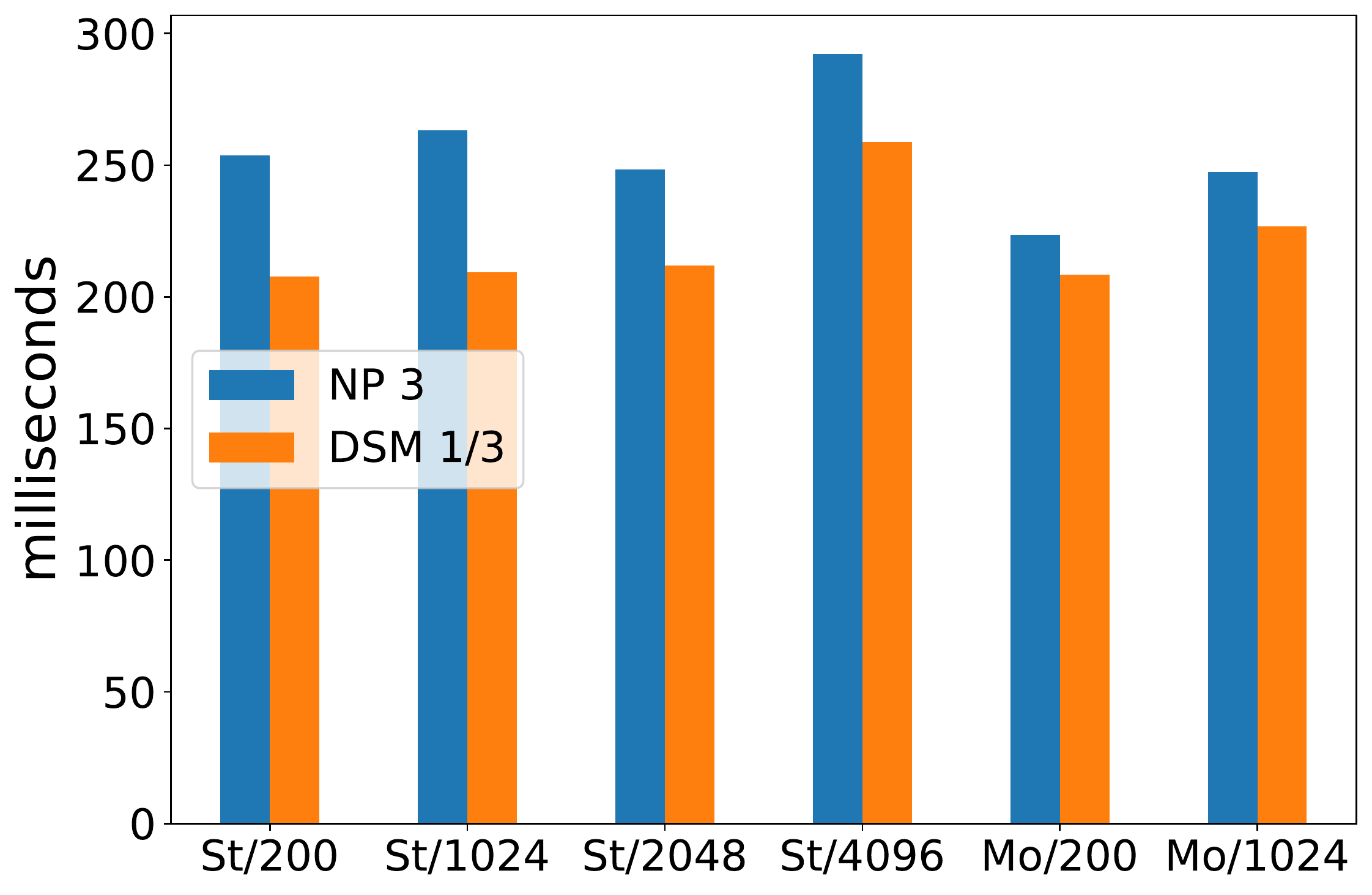}
         \caption{TCP - RTT}
         \label{fig:tcp_rtt}
     \end{subfigure}
      \hfill
     \begin{subfigure}[b]{0.45\textwidth}
         \centering
         \includegraphics[width=6cm, height=4.0cm]{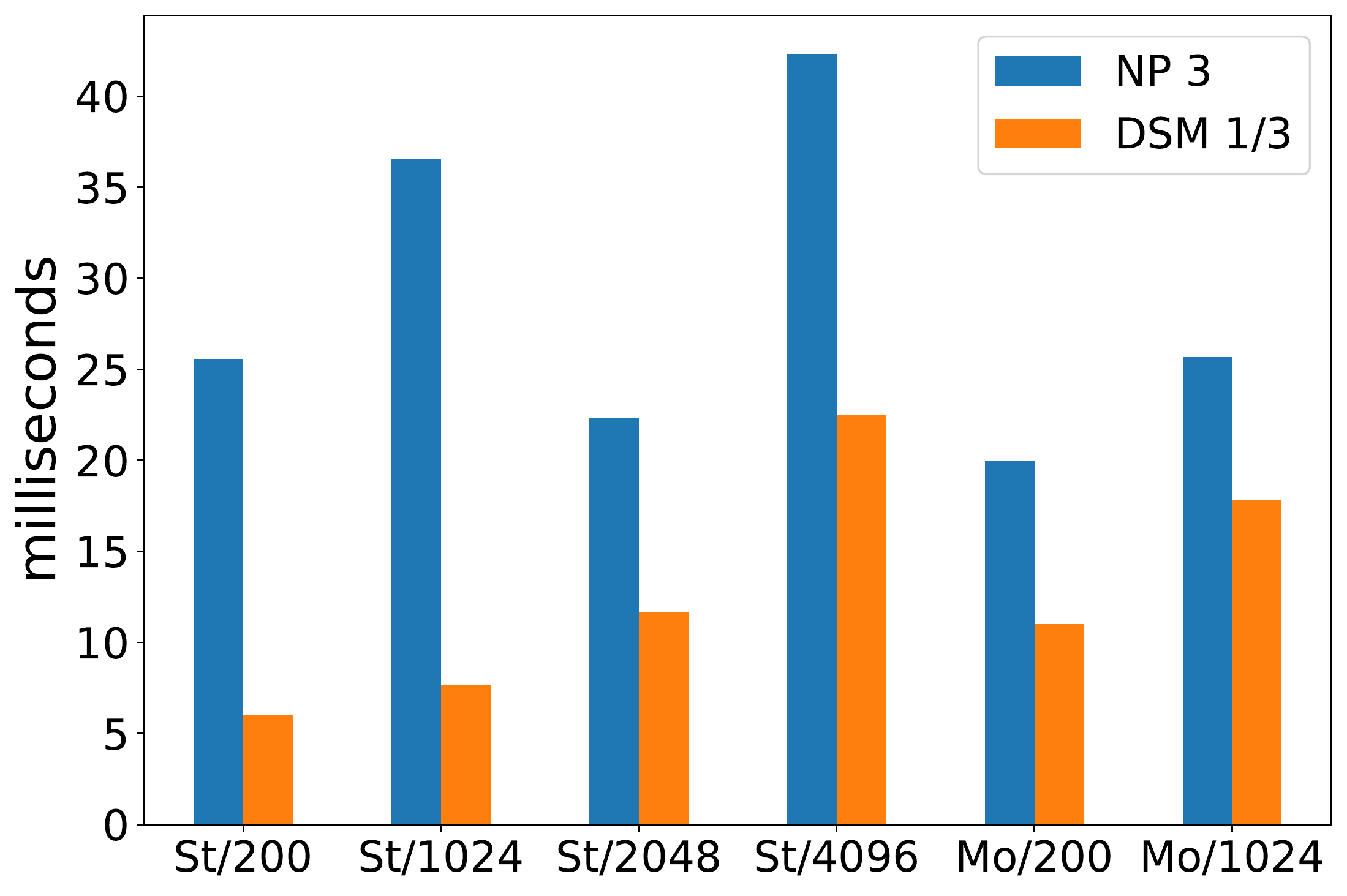}
         \caption{TCP - Jitter}
         \label{fig:tcp_jitter}
     \end{subfigure}
      \hfill
          \begin{subfigure}[b]{0.45\textwidth}
         \centering
         \includegraphics[width=6cm, height=4.0cm]{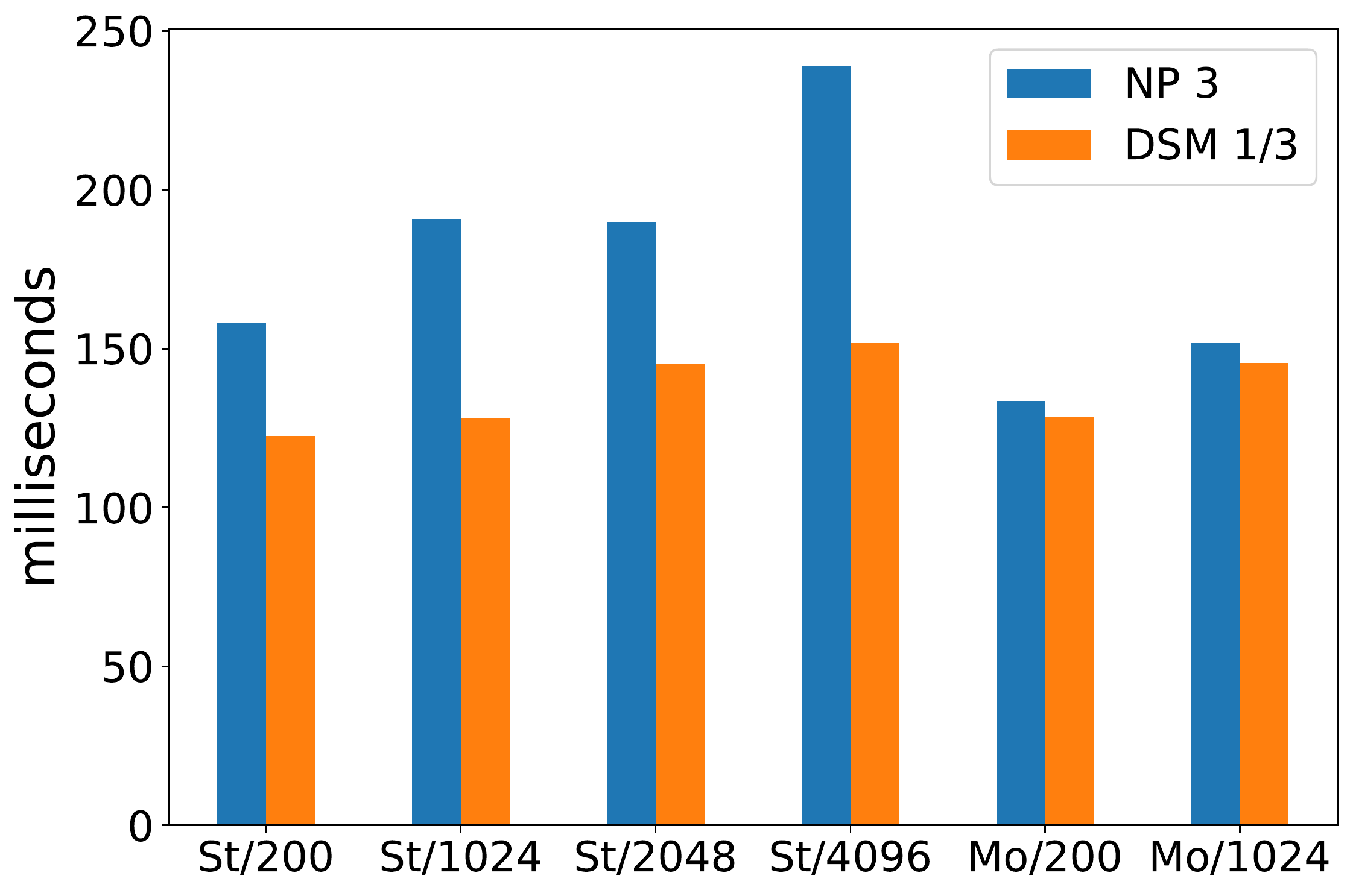}
         \caption{UDP - RTT}
         \label{fig:udp_rtt}
     \end{subfigure}
      \hfill
     \begin{subfigure}[b]{0.45\textwidth}
         \centering
         \includegraphics[width=6cm, height=4.0cm]{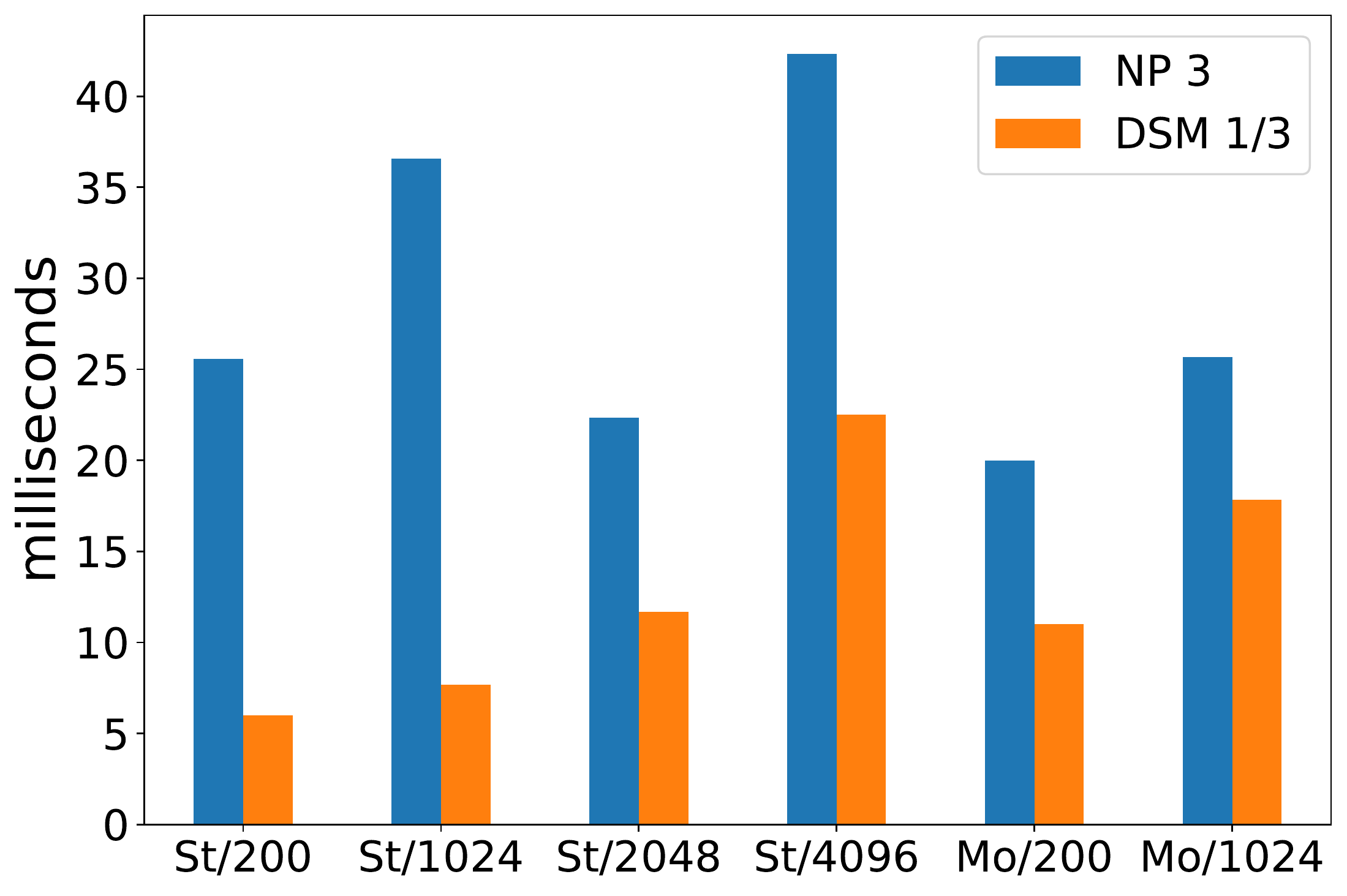}
         \caption{UDP - Jitter}
         \label{fig:udp_jitter}
     \end{subfigure}
        \caption{Day 3/4: Median Latency (RTT) and Jitter achieved by NP 3 vs DSM-MoC 1/3 for different TCP \& UDP packet sizes in stationary (st) and mobile (mo) scenarios. For simplicity, we compared only with NP 3 because we have already established from Day 1 that NP 3 had the best network performance on the route. Three clear insights emerge. First, a DSM-MoC implementation consistently achieved better RTT and jitter for all packet sizes and in stationary and mobile environments. Second, UDP improvement was much higher, suggesting that a DSM-MoC approach is advantageous in use cases with UDP-reliant applications. Third, RTT measured on the Raspberry Pi is significantly worse than RTT measured on smartphones on Day 1, further highlighting the difference between hypothetical scenarios and system-level implementation.}
        \label{fig_tcp_udp}
\end{figure*}

Regardless of whether we are using DSM-MoC or a single network operator only, our assumption for Day 3/4 experiments in a suburban stationary location with few or no changes in base station connections is that there would be less impact on TCP/UDP performance as there would be fewer cases of abruptly terminated TCP/UDP connections. Figure \ref{fig_tcp_udp} confirms that this assumption is correct. The opposite is also true - in a mobile test scenario, TCP/UDP performance deteriorates more than in a stationary test scenario. In fact, \cite{xiao2014tcp} reports that the impact of switching TCP sessions can be as large as 10s. We recognize that this is applicable to DSM-MoC and the impact of switching networks, in addition to switching cell sites,  will impact the relative performance advantages expected. Nonetheless, even for the mobile environment with cell handovers and inter-operator changes, Figure \ref{fig_tcp_udp} is still clear that DSM-MoC providers a superior performance than a single operator. We observe that, for all packet sizes, RTT for data transfer on both TCP and UDP on DSM-MoC achieve $>$10\% improvements over NP 3 in a stationary environment and single digit improvements in a mobile environment. For jitter, the improvements are bigger but again, with better improvements from switching in a stationary scenario.

\begin{figure}[t]
\centering
\includegraphics[width=0.75\columnwidth]{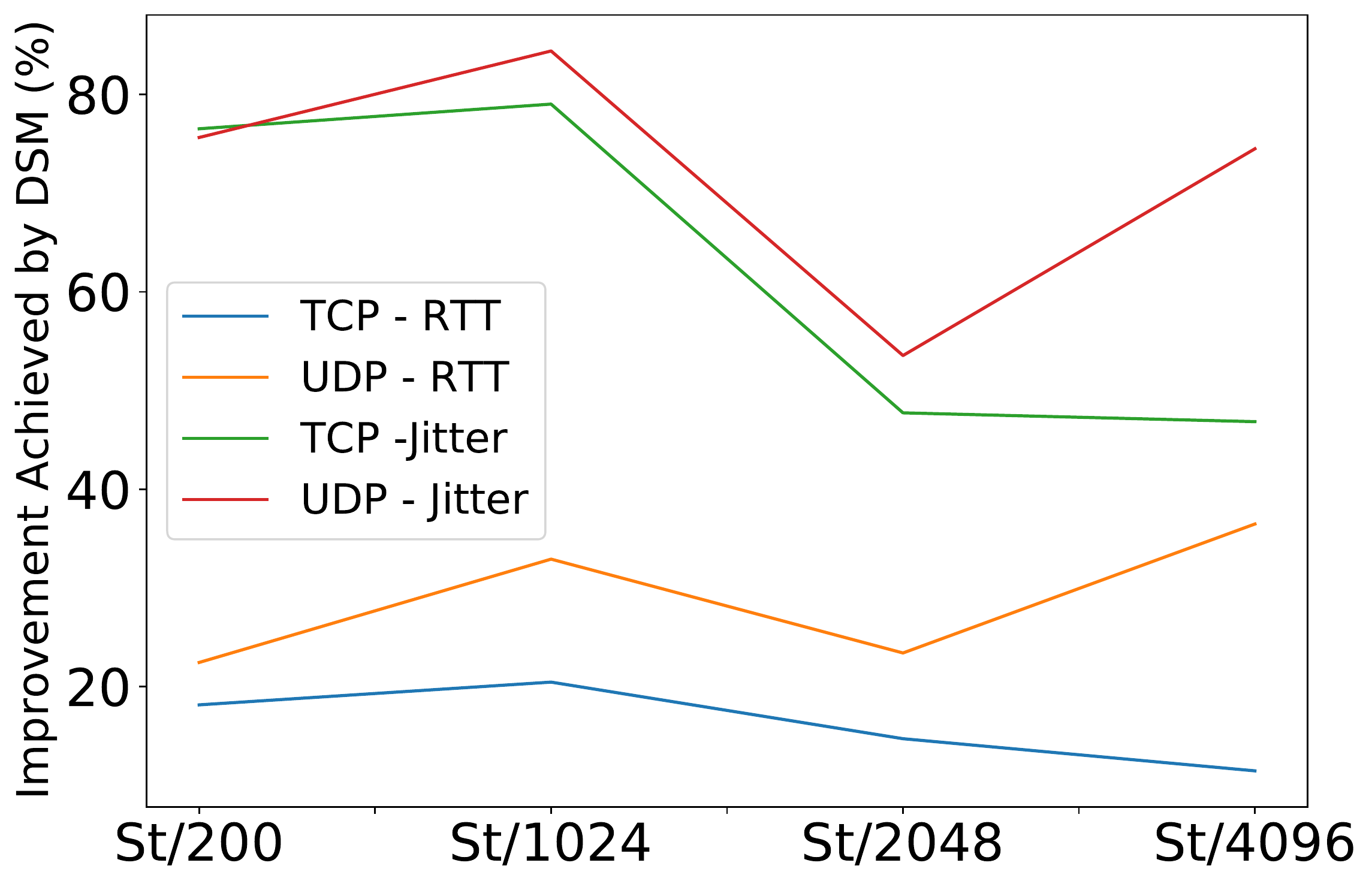}
\caption{Day 4: Stationary DSM-MoC benefits was highest for TCP/UDP packet sizes of 1024 bytes (except outlier for 4096 bytes on UDP) contrary to expectations that 200 byte packets will witness the best improvements}
\label{fig_1024b}
\end{figure}

\subsubsection{Smaller packet sizes did not achieve best improvement}

Another assumption we had for Day 3/4 experiments was that smaller packet sizes will be less susceptible to disruption as the TCP/UDP packets are small enough to be transmitted in the short time windows. However, our results show that we were wrong as the benefits of switching do not favour the lowest packet size of 200 bytes. Instead, Figure \ref{fig_1024b} shows that measurements with 1024 bytes packet sizes had the best performance for both TCP and UDP for RTT and jitter. Our hypothesis is that as 1024 bytes is closer to the typical packet sizes of ~1500 bytes, it could be that the internet infrastructure has been optimised for the typical packet sizes.

\subsubsection{System implementation differs from smartphone implementation}

An interesting insight from Figure \ref{fig_tcp_udp} is that the RTT and jitter measured on the Pi in the CC on Day 3/4 is significantly worse than the RTT and jitter measured on smartphones on Day 1 on the same route. This insight highlights a major difference between our work and previous works which have focused on multi-connectivity on smartphones. In our system implementation, the smartphone acts as the radio unit while the Raspberry Pis and the Netgear router are the main computing units running the application. The additional footprint and distance means that, compared to a smartphone-only setup, the data traffic traverses more network nodes and frontiers which would explain the longer RTT and jitter.

\section{Discussions, Implications \& Recommendations}

\label{discussion}

\subsection{DSM-MoC delivers superior performance}

\begin{figure}[t]
\centering
\includegraphics[width=6.5cm, height=6.5cm]{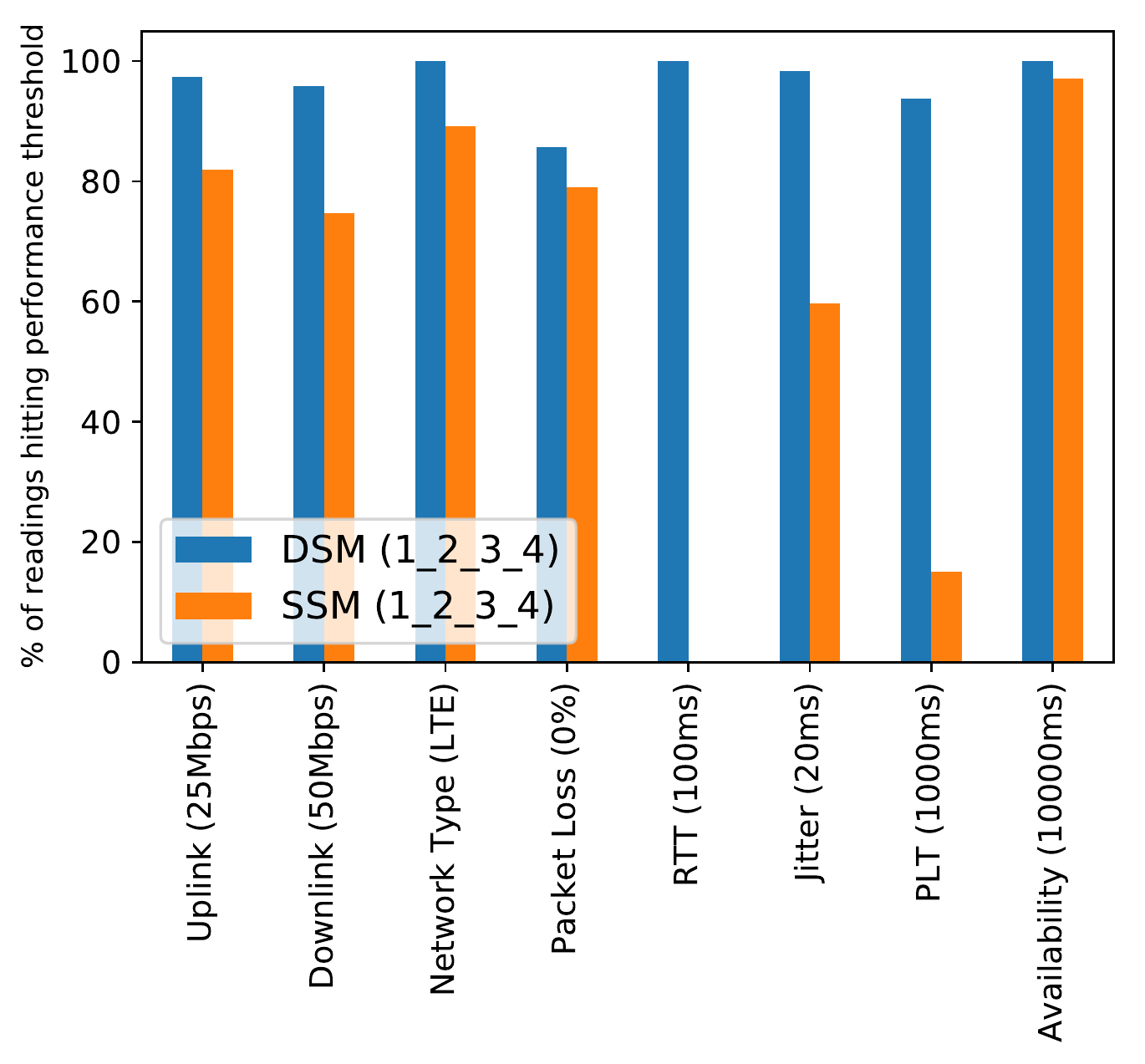}
\caption{Day 1: Comparison of SSM vs DSM performance showing that SSM significantly under-performs a similar DSM}
\label{fig:ssm_dsm}
\end{figure}

\begin{figure}[t]
\centering
\includegraphics[width=6cm, height=6.0cm]{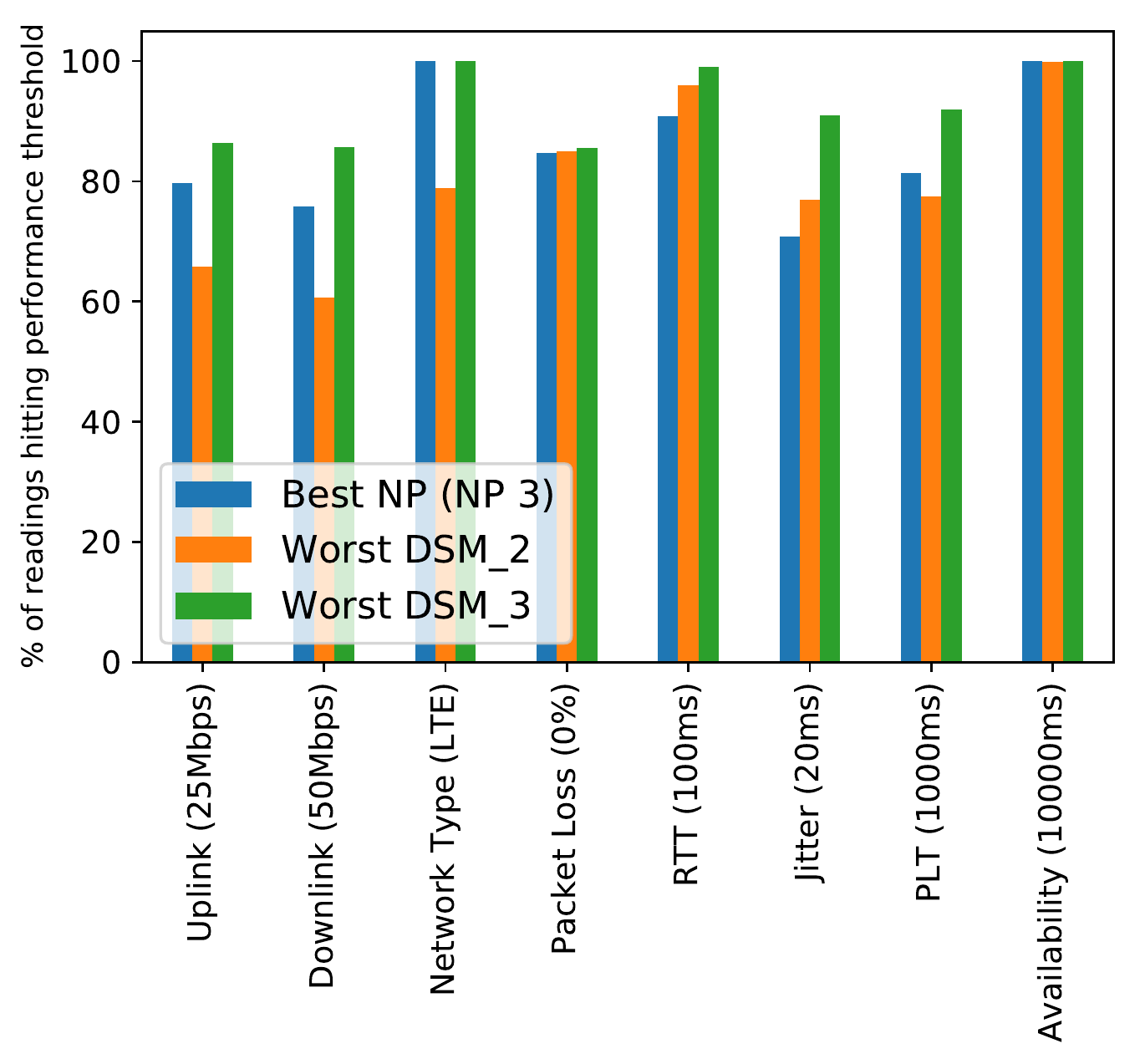}
\caption{Day 1: Comparison of best NP vs the worst 2-network \& 3-network DSM to understand optimal fallback options}
\label{fig:2_3_dsm}
\end{figure}

Whether from Day 1 hypothetical scenario or Day 2 actual measurement data, it is clear that DSM-MoC delivers superior performance. For a 4-network redundancy setup, Figure \ref{fig:ssm_dsm} shows that across all parameters, a hypothetical DSM-MoC setup will deliver a superior performance (based on Day 1 measurements). In fact, based on Day 2/3/4 results, even the 2-network DSM-MoC, as long as it includes NP 3, will deliver superior performance compared to the 4-network SSM-MoC. Yet, digging deeper, we see from Figure \ref{fig:2_3_dsm} for Day 1 that the worst 2-network DSM-MoC would have under-performed the best performing single network (NP 3) for uplink/downlink speeds and network type. In contrast, the worst 3-network fallback option will consistently outperform NP 3. 

While this observation can not be generalised, it suggests that using only a 2-network DSM-MoC cannot always provide assurances of superior performance, especially if those two networks are the worst performing. As such, rather than rely on the two worst networks for QoS assurance, a CC manufacturer is better off striking an exclusive deal with the best-quality network provider. In practice, and before factoring in network sharing, most countries have at least three network providers (UK has four), making it feasible to deliver DSM-MoC.

\begin{figure}{}
\centering
\includegraphics[width=6cm, height=4.5cm]{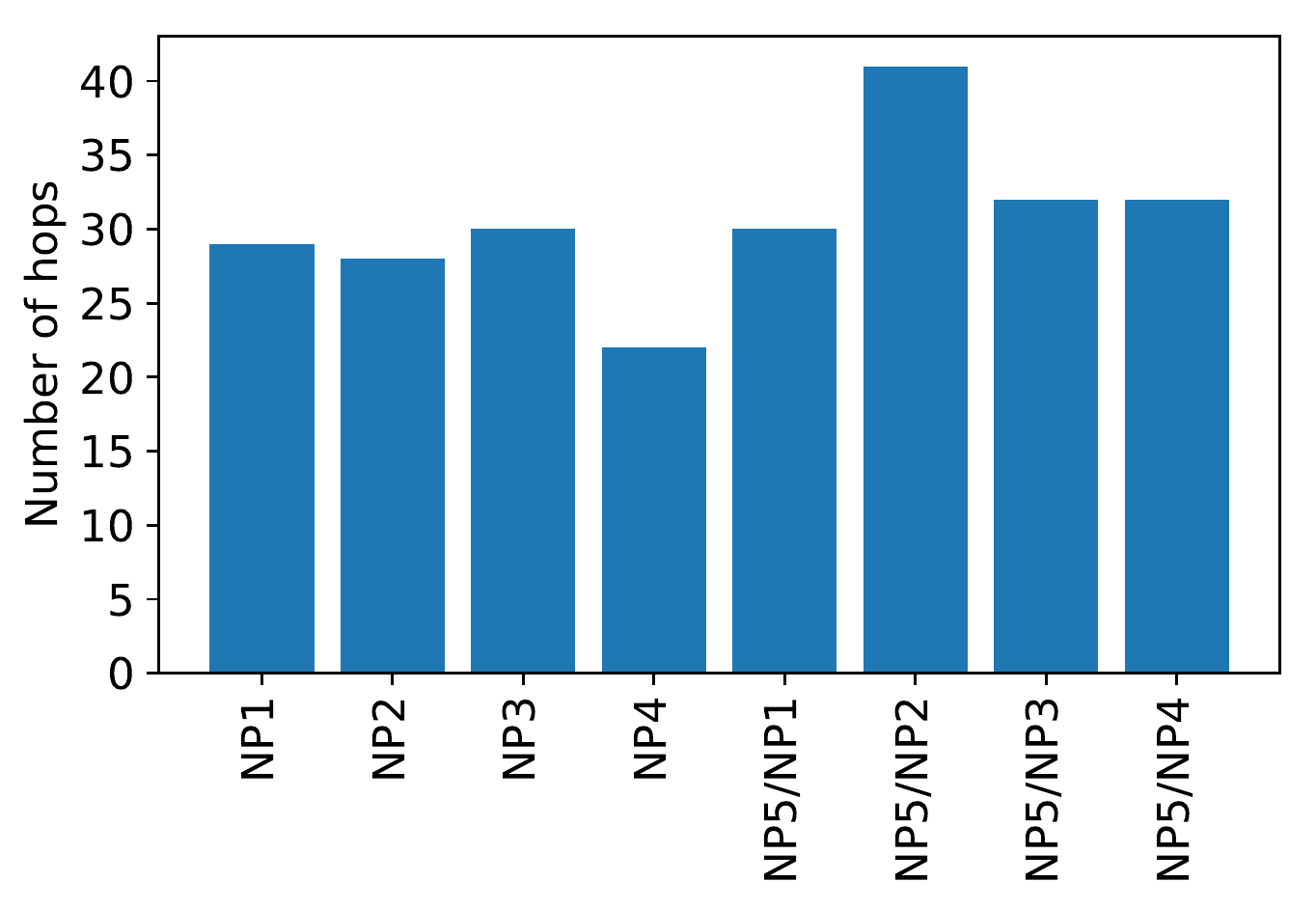}
\caption{No of traceroute hops to 3.8.114.122, showing that SSM-MoC takes a much longer route}
\label{fig:hops}
\end{figure}

\subsection{SSM-MoC's network architecture creates challenges}
In our setup, we note specifically that the SSM-MoC operator promotes its service as being able to optimise on coverage, performance and price. In contrast, our data suggests that it has, instead, optimised mostly for speed to the detriment of other performance indicators. The implication is that the performance we have seen for SSM-MoC contradicts the expectation of using such setup (whether via an MVNO or national roaming) and the justification for its hefty fees.  

Our investigation suggests that the relative under-performance of SSM-MoC is an intrinsic outcome of its architectural design. The SSM-MoC provider makes it clear that it routes cellular traffic through its own cloud-based mobile core network, presumably with its own packet gateway (PGW). We confirm this longer route in Figure \ref{fig:hops} by noting the number of traceroute hops to reach our AWS server. By design, we would expect this setup to negatively impact latency-related measurements (e.g. RTT, jitter and PLT) as the route from our measurement device to the application server is elongated. However, on the purely network-dependent parameters (e.g. downlink/uplink speeds, network type), our expectation was that SSM-MoC should be outperforming NP 1 - 4 as it is capable of switching host NPs to achieve better performance. This was only the case for uplink/downlink speed where SSM-MoC's 25.5Mbps and 50.7Mbps median was the highest. Unfortunately, our analysis of the root cause of this is limited given that SSM-MoC had the same base station ID with NP 2 for only 7\% of readings (and none with NP 1/3/4).

\subsection{Operational complexities for DSM-MoC are surmountable}
In our Day 2/3/4 experiments, we have successfully demonstrated that DSM-MoC can be implemented in a test environment, albeit with teething problems that can be resolved to enable its deployment in a production-grade, commercial environment. For example, we note that there are commercial offerings, using an AI engine in the cloud on the DSM-MoC principle, for smart ambulances (e.g Juniper's Contrail SD-WAN solution demoed at Mobile World Congress 2019\footnote{https://blogs.juniper.net/en-us/industry-solutions-and-trends/smart-ambulance-demo-at-mwc-showcases-critical-5g-sd-wan-use-case}). 

\subsection{Multi-radio connectivity as default for CCs}

A major hindrance to the use of DSM-MoC is on how many radios to have in a CC and how many should be switched on simultaneously. This question determines the power efficiency of the connectivity unit and, ultimately, the overall cost of the system. Despite these concerns, and while a full examination of the energy efficiency are out of scope in our work, our expectation is that CCs should have at least two radios switched on at the same time in keeping with the expectation for redundancy as contained in ISO 26262 (``Road vehicles - Functional Safety") and its master guide, IEC61508 (``Functional Safety of Electrical/Electronic/Programmable Electronic Safety-related Systems")\cite{negi2014redundancy}. We argue that this should be a prerequisite for any vehicle operating on a fully autonomous basis. A multi-radio connectivity provides a fall-back, redundant option and also makes it possible to split the data packets across several radios or to send all the data packets simultaneously through all the radios.

\subsection{CCs have computational power to manage DSM-MoC}
\begin{figure}[t]
\centering
\includegraphics[width=8.5cm, height=5.5cm]{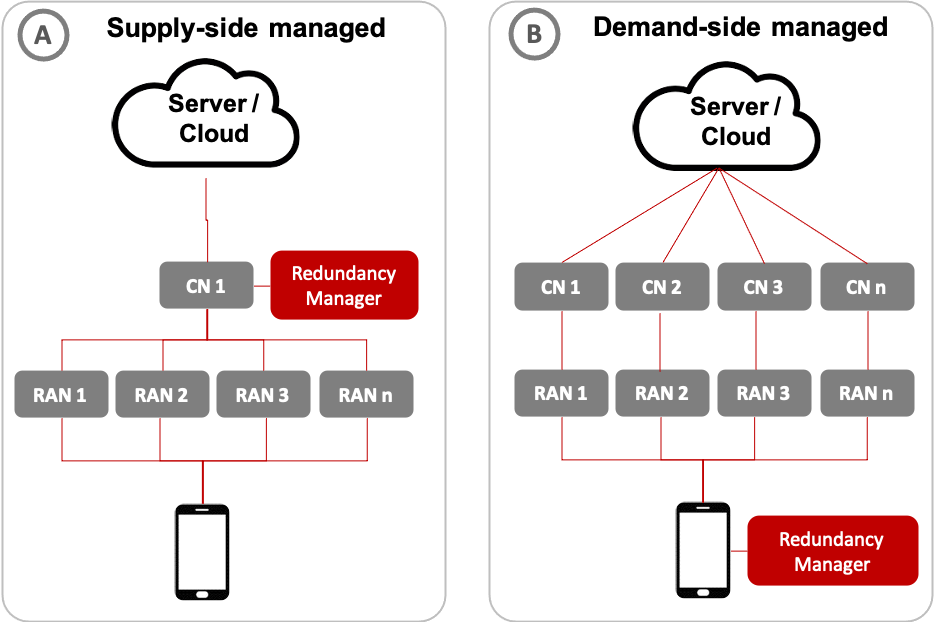}
\caption{Where is redundancy managed? [A] SSM-MoC - Primary network operator manages redundancy at the core network (CN) and makes decisions on which RAN to use. [B] DSM-MoC - User manages redundancy at the local level and selects RAN to optimise performance requirements. Modern cars have enough computing power for DSM-MoC}
\label{dsm_rac}
\end{figure}

In a multi-radio setup, the question of who makes the decision on which radio to use is at the heart of whether the system is SSM or DSM. Figure \ref{dsm_rac} illustrates this dilemma on whether to repatriate the decision making to the nodes/CCs or not. In this work, we argue that cars, as `datacentres on wheels', have the requisite computing power and AI capability to automate such decisions. This explains why driverless cars are increasingly able to make their decisions onboard instead of relying on a central coordinator\cite{mckinsey2021}, an approach in contrast to the industry default where the design philosophy of the cellular network has encouraged dumb nodes and intelligent core. This philosophy explains SSM-MoC, National Roaming and new initiatives such as Juniper's Contrail SD-WAN solution. Under this approach, if a user decides not to hand over decision making to a centralised authority,  then the user with multi-SIM devices have to manage the multi-connectivity themselves, including resetting/restarting their devices where necessary. With our manual setup, we learnt how difficult this can be from our Day 2/3/4 measurements. Despite months of setup and test runs, the system still misbehaved on the day and required multiple manual restarts to ensure seamless operations.

\subsection{Cost constraints discourage DSM-MoC}

Based on current industry structure, DSM-MoC is likely unaffordable in its current setup requiring multiple radios. In a car industry with razor-thin margins, the cost of supporting multiple connectivity options is almost a non-starter. This probably explains why most CC deals are exclusive with a single network provider. McKinsey forecasts that connectivity will present a revenue potential of between \$130 to \$210 for basic connectivity and \$400 to \$610 for advanced connectivity; annual cost savings are in the range of \$100 to \$170 and \$120 to \$210 per vehicle, respectively\cite{mckinsey2021}. Putting them together, the total value from connectivity per car will range from \$230 to \$820, giving an average of \$530 per annum or \$44.2 per month, a value that is similar to the average revenue per user for a typical  mobile phone user in the US. Our industry view is that IoT monthly ARPU is significantly less than smartphone ARPU. However, even if IoT can achieve the same \$44 per month, it is still a huge challenge to profitably divide the \$44 across multiple network providers.

\subsection{Policy guidance is imperative}

In a scenario where safety is the goal, ultimately, the DSM-MoC vs SSM-MoC trade-off will come down to policy guidance. We note that our measurements, on mature 4G networks in a country with over 10 years of 4G availability,  highlight the inadequacy of a single cellular connectivity to meet stringent performance requirements. While 5G will bring improvements to reliability, we are not convinced that it will suddenly make any single network adequate. Performance studies on early 5G networks in China\cite{xu2020understanding}, UK\cite{obiodu20215g} and USA\cite{narayanan2020first} do not yet suggest any significant performance improvement above 4G nor would 5G coverage become extensive any time soon to cover all roads. Given that the industry expectation is for 5G \& 4G to coexist well into the 2030s\cite{obiodu20195g}, we argue that our proposals for DSM-MoC are applicable now and deserve attention from policymakers, commercial executives and technical planners today. 

We recognize existing steps towards improved reliability for CCs. Already, there are rules in place to support fallback and roaming for selected services (e.g. eCall in Europe\cite{oorni2017vehicle}). 3GPP has also made a start by reserving QCI/5QI 3 and 79 for V2X messages or by proposing special network slices for V2X\cite{campolo20175g}. While these mechanisms can improve the reliability, they are unlikely to provide as much geographic and performance reliability as using redundant connectivity options, especially for safety-critical V2X use cases (e.g platooning). Besides, as argued in  \cite{obiodu2018towards}, technical specification without an enabling commercial and policy framework is unlikely to be actualised. 

In the spirit of technology neutral policy making, a concerted push by regulators and policymakers for stringent connectivity KPIs for all cars will nudge the industry towards developing commercial \& technical models for DSM-MoC.

\section{Conclusion}
\label{sec:conclusion}

Connected Cars (CCs) are becoming increasingly common on the roads and the expectation is that they will be served by ultra reliable connectivity solutions to assure safety of drivers and passengers. Our work has investigated how to improve connectivity reliability and provide QoS assurance for CCs using a Demand Side Managed Multi Operator Connectivity (DSM-MoC) setup. Unlike other works that are mostly for price/quality arbitrage on smartphones, the novelty of our work is that we focus on assuring reliability for safety (i.e. safety arbitrage) on a system-level implementation in real-time and real-life on a CC. Based on actual measurements on 800 kilometers of major and minor roads in South East England, over four days, and on the networks of the four UK mobile providers and a global/Universal SIM card provider, we show that DSM-MoC delivers superior performance than reliance on a single network operator. 

On Day 1, we focused on measurements on the four UK networks plus the Universal SIM and using the observations to determine hypothetical effectiveness of DSM-MoC. On Day 2, we built a system-level implementation to actually switch the network connections in real time. For Day 3/4, we measured TCP/UDP performance in a mobile vs stationary scenario while switching networks. 

Based on Day 1, our findings show that a DSM-MoC implementation on a CC will deliver superior performance of up to 28 percentage points in a hypothetical scenario. We also show that the Universal SIM which implements a Supply Side Managed Multi Operator Connectivity (SSM-MoC) has an inferior performance - up to 4.8x longer page load times - while being significantly more expensive (e.g 95x more per 1GB). From Day 2 results, we show that in an actual real life system implementation of network switching the superior performance is at least 12\% for page load times. From Day 3/4 results, we show that despite concerns about dropped TCP/UDP packets during network switching, the superiority of DSM-MoC over a single network is confirmed. We also show that this superiority is more pronounced for UDP packets.

\section{Acknowledgment}
\label{sec:acknowledgment}
 The work is supported by EU H2020 grant agreements No 871793 (Accordion), No 101017109 (Daemon) and No 101016509 (Charity). Special thanks to Alex Geoman (Fujitsu, UK) for the support in building the test-bed for Day 2/3/4.

\bibliographystyle{unsrt}
\begin{small}
\bibliography{biblio}
\end{small}

\begin{IEEEbiography}[{\includegraphics[width=1in,height=1.0in,clip,keepaspectratio]{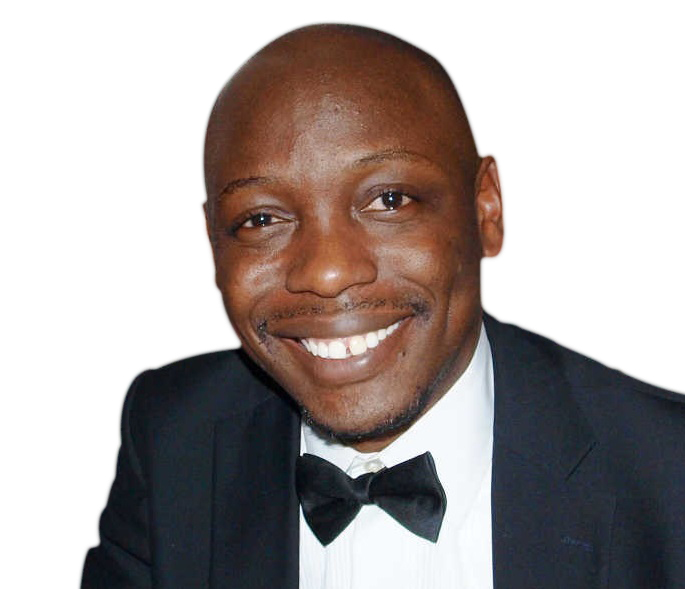}}]{Emeka Obiodu} (chukwuemeka.obiodu@kcl.ac.uk) is a PhD student in the Department of Engineering at King’s College London where his research focuses on techno-economic dynamics of differentiated services in the 5G era. Previously, he was Strategy Director at GSMA, and edited the GSMA's 5G handbook. He has a BEng from Federal University of Technology Owerri, an MSc in Telecommunications from Queen Mary University of London, and an MBA from the Warwick Business School. 

\end{IEEEbiography} 
\vspace{-16 mm}

\begin{IEEEbiography}[{\includegraphics[width=1in,height=1.0in,clip,keepaspectratio]{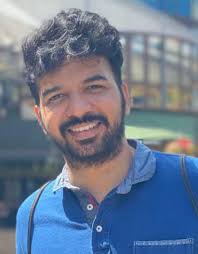}}]{Aravindh Raman} (aravindh.raman@telefonica.com) received a Ph.D. degree from the Department of Informatics, King’s College London. He is a Researcher with Telefonica I+D in Spain and has previously being with IIT Delhi and Tata Institute of Fundamental Research, Mumbai, India. His research interests are on future Internet infrastructure, network measurements and building next-generation content delivery architectures. 

\end{IEEEbiography} 
\vspace{-16 mm}

\begin{IEEEbiography}[{\includegraphics[width=1in,height=1.0in,clip,keepaspectratio]{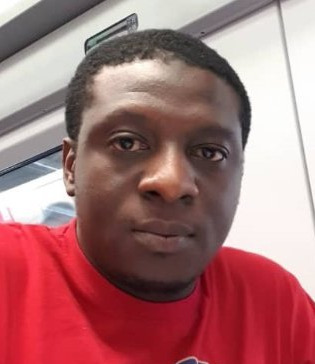}}]{Abdullahi Kutiriko Abubakar}  (a.abubakar@surrey.ac.uk) is a PhD research student in the Department of Computer Science, University of Surrey. His research interests span IoT, Community Networks, Long Range Wide Area Networks and Edge Computing. Before now, He was a research student at King’s College London under the supervision of Prof. Nishanth Sastry. 

\end{IEEEbiography} 
\vspace{-16 mm}

\begin{IEEEbiography}[{\includegraphics[width=1in,height=1.0in,clip,keepaspectratio]{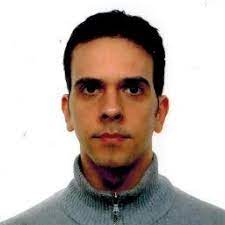}}]{Simone Mangiante} (simone.mangiante@vodafone.com) received a PhD in computer networks from the University of Genoa, Italy, in 2013. He is a Research \& Standards Specialist at Vodafone UK, and was previously with Dell EMC Research Ireland. His research interests are focused on computer networks, software defined networking, cloud architecture, and Internet of Things. He has been involved in European projects focusing on SDN (FP7 Marie Curie SOLAS) and network transport (H2020 NEAT). 

\end{IEEEbiography} 
\vspace{-16 mm}

\begin{IEEEbiography}[{\includegraphics[width=1in,height=1in,clip,keepaspectratio]{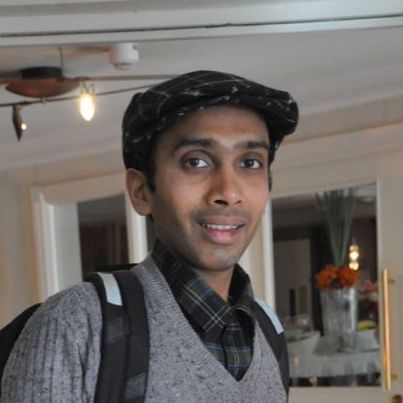}}]{Nishanth Sastry} (n.sastry@surrey.ac.uk) is a Professor of Computer Science at University of Surrey, UK. Prior to this he was at the Dept. of Informatics at King's College London. He holds a PhD in Computer Science from the University of Cambridge. His current research interests include computer and social networks, computational social science, and data analytics aspects of these two areas. He has been a visiting researcher at the Alan Turing Institute and MIT. Nishanth is a member of the ACM and a Senior Member of the IEEE.
\end{IEEEbiography}
\vspace{-16 mm}

\begin{IEEEbiography}[{\includegraphics[width=1in,height=1.0in,clip,keepaspectratio]{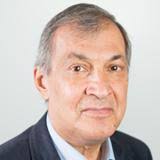}}]{A. Hamid Aghvami} (hamid.aghvami@kcl.ac.uk) (M’89–SM’91–F’05, Life Fellow, IEEE) was a Visiting Professor with NTT Radio Communication Systems Laboratories in 1990, a Senior Research Fellow with BT Laboratories from 1998 to 1999, and an Executive Advisor with Wireless Facilities Inc., USA, from 1996 to 2002. He is currently a Professor of telecommunications engineering with King’s College London. He has authored or co-authored over 600 technical papers and given invited talks and courses worldwide on various aspects of personal and mobile radio communications. He was a member of the Board of Governors of the IEEE Communications Society from 2001 to 2003, and a Distinguished Lecturer of the IEEE Communications Society from 2004 to 2007. In 2009, he was a recipient of the Fellowship of the Wireless World Research Forum in recognition of his personal contributions to the wireless world. He has been a member, a Chairman, and a Vice-Chairman of the technical program and organizing committees of a large number of international conferences. He is a Fellow of the Royal Academy of Engineering and a Fellow of the IET. 

\end{IEEEbiography} 
\vspace{-16 mm}

\end{document}